\documentclass[a4paper,11pt]{article}
\pdfoutput=1 	
\usepackage{jheppub} 	
\usepackage{verbatim}
\usepackage{graphicx}	
\usepackage{dcolumn}	
\usepackage{bm}		    
\usepackage{hyperref}
\usepackage{amsmath}
\usepackage{verbatim}
\usepackage{slashed}
\usepackage{hyperref}
\usepackage{mathtools}
\usepackage{tabularx}
\usepackage{xfrac}

\title{\boldmath  Comparable Dark Matter and Baryon energy densities from Dark Grand Unification}

\author{Yi Chung}

\affiliation{
Max-Planck-Institut für Kernphysik, Saupfercheckweg 1, 69117 Heidelberg, Germany
}

\emailAdd{yi.chung@mpi-hd.mpg.de}

\abstract{We investigate a theory of $SU(9)$ dark grand unification, where dark matter consists of asymmetric dark baryons from the $Sp(4)_D$ dark QCD sector. By unifying the dark color gauge group with the Standard Model gauge group, the asymmetry generation in both sectors originates from a common process that preserves a $U(1)_{D-(B-L)}$ symmetry, resulting in comparable number densities. Furthermore, thanks to dark grand unification, the $Sp(4)_D$ dark QCD sector shares a similar matter content with the QCD sector, leading to comparable running of the gauge couplings from the scale as high as $10^{15}$ GeV. This predicts a dark color confinement scale and thus dark baryon masses around the GeV scale, comparable to visible baryon masses. Together with the similar number densities, the model provides an explanation for the observed similarity between the energy densities of dark matter and baryons, $\rho_D \approx 5\,\rho_B$. The model also features some novel phenomenology, including a GeV-scale flavored dark QCD sector with diquark dark baryons and light dark mesons. The interaction between the dark sector and the visible sector occurs via a new $Z'$ boson with a mass of $\mathcal{O}(10)$ TeV, which could be searched for at future hadron colliders. We also briefly discuss an $SU(8)$ dark grand unified theory featuring an $SU(3)_D$ dark QCD sector.}

\begin{document}
\maketitle
\flushbottom

\section{Motivation}


The Standard Model (SM) of particle physics provides an excellent description of all known elementary particles and their interactions. Nonetheless, a number of open questions remain unanswered, especially those revealed by astronomical and cosmological observations, including the existence of dark matter, dark energy, and the origin of the Universe. Among these mysteries, understanding the nature of dark matter remains one of the most compelling challenges in particle physics.


So far, all the evidence regarding the existence of dark matter comes from its gravitational effects and the only precise quantitative measurement of dark matter is its energy density, $\Omega_c\equiv \rho_D/\rho_{\rm crit}\,$, which can be determined by its impact on the evolution of the Universe. The latest measurement of the cosmic microwave background (CMB) anisotropies from the Planck collaboration \cite{Planck:2018vyg} (considering TT, TE, EE + lowE + lensing) gives
\begin{align}
\Omega_c h^2 = 0.1200 \pm 0.0012 \qquad \text{where} \quad ~h=H_0/100 ~\text{(km/s/Mpc)}~.
\end{align}
The energy density alone provides only limited information about the nature of dark matter. However, if one also takes into account the baryon energy density, $\Omega_b h^2 = 0.0224 \pm 0.0001$, from the same measurement, a mysterious $\mathcal{O}(1)$ ratio $\Omega_c / \Omega_b = \rho_D/ \rho_B\approx 5$ emerges, known as the Dark Matter-Baryon Coincidence. Such a coincidence is especially surprising when one notices that the ratios among the SM non-relativistic species, such as proton-to-electron energy density ratio, $\rho_p / \rho_e \approx 1800$, or proton-to-neutron energy density ratio, $\rho_p / \rho_n \approx 7$, are both greater than $\rho_D / \rho_B \approx 5$. In particular, the $\mathcal{O}(1)$ proton-to-neutron energy density ratio is associated with the well-known $SU(2)$ flavor symmetry. Therefore, this coincidence might imply a deeper connection between the two sectors and offers another route to probe the microscopic nature of dark matter.


If the dark matter energy density originates from a mechanism entirely unrelated to baryogenesis, the $\mathcal{O}(1)$ ratio between the two densities would require extreme fine-tuning in the parameter space.\footnote{The coincidence can also be explained if new dynamics exist that directly balance the energy densities of the two sectors, see \cite{Brzeminski:2023wza,Banerjee:2024xhn} for example.} Therefore, the dark sector must share some similarities with the QCD sector. To analyze the coincidence problem, we first separate the energy density $\rho$ of non-relativistic particles $i$ into "number density $n$ $\times$ particle mass $m$", i.e. $\rho_i=n_i\times m_i$. Start with the number density $n_i$. As we already know, the observed baryon number density arises from the small asymmetry between baryon and anti-baryon through the process of baryogenesis. By analogy, dark matter is expected to behave similarly. The motivation leads to the idea of Asymmetric Dark Matter (ADM) \cite{Spergel:1984re,Nussinov:1985xr,Gelmini:1986zz,Barr:1990ca,Barr:1991qn,Kaplan:1991ah,Gudnason:2006ug,Gudnason:2006yj,Kaplan:2009ag,Davoudiasl:2012uw,Petraki:2013wwa,Zurek:2013wia}, where a conserved $U(1)$ global symmetry is introduced to ensure the number densities $n_D\approx n_B$ between the two sectors.


However, this mechanism alone is insufficient.  To fully address the coincidence problem, the particle masses in the two sectors must also be correlated. Since the baryon mass is determined by the QCD confinement scale $\Lambda_{\rm QCD}$, a similar structure should exist in the dark sector, which requires a new strong interaction - dark color (DC). The dark matter candidate would then consist of dark baryons with masses determined by the dark color confinement scale $\Lambda_{\rm DC}$. Ensuring that the two confinement scales are of the same order is highly nontrivial, as they are exponentially sensitive to UV parameters. For example, in QCD, the baryon mass is given by
\begin{align}
m_B\approx \Lambda_{\rm QCD} \approx 
 e^{{-2\pi}/({b_{s} \,\alpha_{s,0}})}\cdot M_{\text{Planck}}~,
\end{align}
where $b_{s}$ is the coefficient of the one-loop beta function in QCD, which is determined by the gauge group and the matter content that is charged under it, and $\alpha_{s,0}$ denotes the coupling constant of $SU(3)_C$ at the Planck scale.


Several approaches have been explored, including symmetry based solutions (mostly via a $\mathbb{Z}_2$ discrete symmetry) \cite{Hodges:1993yb,Berezhiani:1995am,Foot:2003jt,An:2009vq,Farina:2015uea,GarciaGarcia:2015pnn,Lonsdale:2018xwd,Lonsdale:2014wwa,Lonsdale:2014yua,Ibe:2019ena,Rosa:2022sym,Bodas:2024idn},
which replicate the SM in a mirror dark sector to ensure similar values of $b$ and $\alpha_{0}$. In \cite{Bai:2013xga,Newstead:2014jva,Ritter:2022opo,Ritter:2024sqv}, the infrared fixed points are used to relate the two confinement scales. In this study, we focus on solutions based on the unification of the QCD and DC gauge groups \cite{Barr:2011cz,Barr:2013tea,Barr:2015lya,Murgui:2021eqf}, which naturally explains the identical coupling constants of QCD and DC in the far UV, as well as the origin of the global $U(1)$ symmetry necessary to realize the ADM mechanism.

However, unification alone does not guarantee a similar running behavior between the two gauge couplings after they have been separated by symmetry breaking. If there is a difference between the two coefficients $b$’s, a low unification scale is required to achieve comparable confinement scales, which then faces strong constraints from existing measurements. In \cite{Murgui:2021eqf} with $SU(2)_D$ dark color group, to achieve similar coefficients, additional color states are introduced to slow down the  $SU(3)_C$ running, aligning it with the relatively slow $SU(2)_D$ running. However, this approach appears somewhat ad hoc. Additionally, the dark matter originates from different multiplets than those of the SM fermions, so that unification remains confined to the gauge sector without extending to the fermion sector.

In this work, we follow the other approach by extending the SM fermion sector. The extension of the fermion multiplets typically includes new SM singlets, which can serve as our dark quarks. Furthermore, accidental global symmetries can naturally arise, which can stabilize the dark matter and ensure comparable number densities $n_D\approx n_B$. The idea was previously mentioned in \cite{Barr:2011cz,Barr:2013tea,Barr:2015lya}, but only as a realization of the ADM mechanism without explaining the comparable masses. In our study, we go further by introducing a dark QCD sector to tackle the complete coincidence problem. Following this construction, unification not only ensures identical gauge couplings at a high scale but also guarantees similar matter content. Based on this framework, we study the dark QCD sector with the $Sp(4)_D$ dark color group, which has the same quadratic Casimir as that of the $SU(3)_C$ SM color group and allows for a desired fermion spectrum. With these two properties, a similar running of couplings between the two sectors can be achieved, leading to comparable confinement scales and, consequently, similar dark and visible baryon masses. The issues associated with the naive choice of the $SU(3)_D$ dark color group will also be discussed.


This paper is organized as follows. We start with the construction of the model, the $SU(9)$ dark grand unification with $Sp(4)_D$ dark QCD sector, in section \ref{sec:Model}, including the choice of the dark color gauge group, the fermion content, and the scalar content. With the Yukawa interaction terms and fermion masses established, we present the overall spectrum of the model. The determination of confinement scales and the $U(1)$ symmetry that are critical for achieving comparable energy densities are discussed in \ref{sec:Spectrum} and \ref{sec:U(1)}. Next, we zoom in on the strongly coupled dark sector. The low energy dark hadron spectrum is presented in section~\ref{sec:DS}. With all the matter content and interactions finalized, we discuss cosmology in section~\ref{sec:COS} and phenomenology in section~\ref{sec:Pheno}, including all kinds of observables and constraints. Section~\ref{sec:Conclusion} contains our conclusions. In Appendix~\ref{App:SU(3)}, we briefly discuss the possibility of constructing a similar model based on $SU(8)$ dark grand unification with an $SU(3)_D$ dark QCD sector. We highlight the differences and discuss some drawbacks of such a model. In Appendix~\ref{App:Twoloop}, we carried out a two-loop calculation and found results consistent with the one-loop analysis, ensuring the reliability of our conclusions.


\section{The Model}\label{sec:Model}

In this section, we will construct our model step by step. The ultimate goal is to achieve similar running and thus confinement scales between DC and QCD. With this priority, we first choose a suitable dark color gauge group. Next, we determine the desired matter content by extending the fermion content of the $SU(5)$ grand unified theory. The scalar sector required for symmetry breaking and fermion mass generations will then be introduced.

\subsection{The choice of dark color gauge group}\label{sec:Sp4}

Start with the determination of the dark color group. To get similar confinement scales between DC and QCD, even with help from the unification of couplings at a high scale, the running of the couplings should also be close. For the coupling constant $g_i$ of the gauge group $G$, the one-loop beta function is given by 
\begin{align}\label{1loop}
\frac{d\,g_i}{d\,\text{ln}\,\mu}=-\frac{1}{16\pi^2}\,b_i\,g_i^3 \quad\text{, where}\quad
b_i=\frac{11}{3}C_2(G)-\frac{2}{3}n_fT(R_f)-\frac{1}{6}n_sT(R_s)~.
\end{align}
The $n_f$ and $n_s$ are the number of active Weyl fermions and real scalars charged under the gauge group $G$ respectively. $C_2(G)$ is the quadratic Casimir for the adjoint representation and $T(R)$ is the Dynkin index. From the coefficient, we can see that the gauge boson contribution dominates. Therefore, we prefer a dark color gauge group with the same $C_2(G)=3$ as QCD. This requirement narrows down the choice of the dark color gauge group to three: $SU(3)$, $Sp(4)$, and $SO(8)$. Other choices are still possible but would either result in a much lower unification scale or require specific matter content to fine-tune the running.

Among the three choices, $SU(3)$ is undoubtedly the most straightforward option, which is also the case in unification models based on a mirror dark sector \cite{Lonsdale:2014wwa,Lonsdale:2014yua,Ibe:2019ena}, where the dark QCD sector is the same as the SM QCD sector. However, to achieve unification with the SM gauge group in a conventional manner, $SU(3)$ does not work well. It raises the issue of quantum anomaly, requiring a more complicated setup, and also has problems getting a desired fermion spectrum, as we will illustrate further in Appendix~\ref{App:SU(3)}. On the other hand, when the dark color group is $Sp(N)$ or $SO(N)$, the issue of quantum anomaly is less severe and the desired fermion spectrum can be achieved. Therefore, in this work, we will focus on the $Sp(4)$ dark QCD sector.


\subsection{Fermion content and $SU(9)$ Dark Grand Unification}

Next, we aim to construct our matter content following the idea of grand unification, beginning with the fermion sector. In principle, the fermion content is automatically determined by the way the SM gauge group is unified and extended. Therefore, our first step is to decide how we unify the SM gauge group. Among various candidates for grand unification, we adopt the well-known Georgi-Glashow $SU(5)$ Grand Unified Theory (GUT) \cite{Georgi:1974sy} as our framework, whose advantages has already be shown in \cite{Barr:2011cz,Barr:2013tea,Barr:2015lya} and will also become clear in subsequent discussions.\footnote{A similar construction has been realized in \cite{Barr:2011cz,Barr:2013tea,Barr:2015lya}, but only as a UV theory of ADM models without a dark confining sector to relate the dark matter and baryon masses. Our attempt can be viewed as an extension of this idea to fully address the coincidence problem. On the other hand, in \cite{Murgui:2021eqf}, the authors also work on the unification of DC and QCD but based on a Pati-Salam-like extension \cite{Pati:1974yy}. However, all the new fermions from the SM extension still carry SM charges and therefore cannot serve as dark quarks. As a result, dark quarks can only arise by introducing new multiplets separate from the SM fermions.}

Under the $SU(5)$ GUT, the fermion content of the SM can be embedded into an antisymmetric two-index representation and an anti-fundamental representation as
\begin{align}
\mathbf{10}=
\begin{pmatrix}
u^c     &  q  \\
-q^T     &  e^c  \\
\end{pmatrix}
~\text{ and }~
\overline{\mathbf{5}}=
\begin{pmatrix}
d^c \\ \ell 
\end{pmatrix}~\text{ for each generation}.
\label{SU(5)F}
\end{align}
Next, we extend the $SU(5)$ GUT gauge group to the $SU(9)$ Dark GUT gauge group in order to include the $Sp(4)_{D}$ dark color group. The fermion content in \eqref{SU(5)F} is extended as
\begin{align}
\mathbf{36}=
\begin{pmatrix}
\chi_a \oplus \chi_0     &   \chi_c   &  \chi_w \\
-\chi_c^T   &   u^c      &  q      \\
-\chi_w^T   &   -q^T     &  e^c    \\
\end{pmatrix}
~\text{ and }~
\overline{\mathbf{9}}=
\begin{pmatrix}
\chi_f \\ d^c \\ \ell 
\end{pmatrix}~\text{ for each generation}.
\label{SU(9)F}
\end{align}
Five new types of fermions, all labeled by $\chi$ but with different subscripts, are introduced, transforming under the gauge group $Sp(4) \times SU(3)_C \times SU(2)_W \times U(1)_Y$ as 
\begin{align}\label{chi}
\chi_a= (5,1,1)_0~,~ \chi_0= (1,1,1)_0~,~\chi_c=(4,3,1)_{-\frac{1}{3}}~,~\chi_w =(4,1,2)_{\frac{1}{2}}~,~ \chi_f= (\bar{4},1,1)_0~.
\end{align}
The subscripts reflect their quantum number, with $a$, $f$, and $0$ denoting antisymmetric, fundamental,\footnote{Under the $Sp(4)$ group, there is no distinction between fundamental and anti-fundamental representations. However, we will still label the $4$ and $\bar{4}$ based on their origin in the $SU(4)$ group to help keep track of symmetry patterns.} and singlet representations under $Sp(4)$, respectively, and with $c$ and $w$ indicating it is charged under the SM $SU(3)_C$ and $SU(2)_W$. From the new fermion content, we can immediately identify SM-singlet fermions that can serve as dark quark candidates, which is an advantage of extending the $SU(5)$ GUT to include the dark color group.

However, it is not the whole story since $SU(9)$ is anomalous with this fermion content. To cancel the anomaly, four additional $\overline{\mathbf{9}}$ anti-fundamental representations are required for each generation, which are written as
\begin{align}
4 \times \overline{\mathbf{9}}=\overline{\mathbf{9}}_4=
\begin{pmatrix}
\psi_f \\ \psi_c \\ \psi_w 
\end{pmatrix}
\label{SU(9)F}~.
\end{align}
Again, there are three new types of fermions, all labeled by $\psi$ but with different subscripts. They transform under the group $SU(4)'\times Sp(4) \times SU(3)_C \times SU(2)_W \times U(1)_Y$ as 
\begin{align}\label{psi}
\psi_f= (\bar{4},\bar{4},1,1)_0~,\quad \psi_c=(\bar{4},1,\bar{3},1)_{\frac{1}{3}}~,\quad \psi_w =(\bar{4},1,1,\bar{2})_{-\frac{1}{2}}~,
\end{align}
where $SU(4)'$ is the global symmetry among the four new anti-nonets. We choose them to transform as $\bar{4}$ for future convenience.

There are also new SM-singlet fermions $\psi_f$ that could play a role in the dark sector. However, exotic fermions $\psi_c$ and $\psi_w$ charged under the SM gauge group also arise. Together with $\chi_c$ and $\chi_w$ in eq. \eqref{chi}, these exotic SM-charged fermions should be heavy enough to escape the current constraints. From the quantum number assignment, it is impossible to write down mass terms without breaking any gauge symmetry. However, notice that their quantum numbers under the SM match perfectly. Meanwhile, $\chi_{c,w}$ are quartets under the $Sp(4)$ gauge group, and $\psi_{c,w}$ are quartets under the $SU(4)'$ global symmetry. This provides us with a novel way to regain the gauge symmetry we want. That is, instead of identifying $Sp(4) \subset SU(9)$ as our dark color group, we further gauge the $Sp(4)'$ subgroup within the $SU(4)'$ global symmetry and identify the diagonal subgroup $Sp(4)_D$ of $Sp(4) \times Sp(4)'$ as our dark color gauge group. Under this new setup, the quantum numbers of fermions in eqs. \eqref{chi} and \eqref{psi} become
\begin{align}\label{newfermion} 
&\chi_a= (5,1,1)_0~,~ \chi_0= (1,1,1)_0~,~ \chi_f= (\bar{4},1,1)_0~,~\chi_c=(4,3,1)_{-\frac{1}{3}}~,~\chi_w =(4,1,2)_{\frac{1}{2}}~,\nonumber\\
&\psi_s= (\bar{10},1,1)_0~,~ \psi_a= (\bar{5},1,1)_0~,~ \psi_0= (1,1,1)_0~,~ \psi_c=(\bar{4},\bar{3},1)_{\frac{1}{3}}~,~ \psi_w =(\bar{4},1,\bar{2})_{-\frac{1}{2}}
\end{align}
under the gauge group $Sp(4)_D \times SU(3)_C \times SU(2)_W \times U(1)_Y$, where the fermions $(\psi_s,\psi_a,\psi_0)$ come from the decomposition of $\psi_f$. We can easily see that there are two pairs of Weyl fermions that can form Dirac fermions, $(\chi_c,\psi_c)$ and $(\chi_w,\psi_w)$. As the mass terms will be introduced later, for simplicity, we rewrite the notation as
\begin{align}
\psi_c^c=\chi_c=(4,3,1)_{-\frac{1}{3}}~,~~\psi_w^c=\chi_w =(4,1,2)_{\frac{1}{2}}~.
\end{align}

Besides the $Sp(4)'$ symmetry, we would like to gauge an additional $U(1)_5$ symmetry, with charges assigned to the different $SU(9)$ multiplets as 
\begin{align}
Q_5({\mathbf{36}})=0,\quad
Q_5(\overline{\mathbf{9}})=1,\quad
Q_5(\overline{\mathbf{9}}_4)=-1/4~.
\end{align}
The $U(1)_5$ symmetry originates from the subgroup of $SU(5)_G$ global symmetry among $\overline{\mathbf{9}}$'s which will be discussed in detail in section \ref{sec:U(1)}. The corresponding $Z'$ boson will play an important role in cosmology and phenomenology.

So far, besides the $SU(9)$ Dark GUT group, we have gauged two additional groups, $Sp(4)' \times U(1)_5$. However, adding new gauge symmetries is not a free lunch, as additional quantum anomalies might arise. For the $Sp(4)'$ gauge symmetry, although there are no perturbative gauge anomalies like $SU(N)$ group due to the lack of complex representations, there is, however, a “non-perturbative anomaly”, called the Witten anomaly \cite{Witten:1982fp}, which requires an even number of Weyl fermions in the fundamental representation. Therefore, for this anomaly to cancel, we add additional fermions
\begin{align}
\chi^c_f= (4,1,1,1)_0
\end{align}
under $Sp(4)'\times Sp(4) \times SU(3)_C \times SU(2)_W \times U(1)_Y$ to each generation. The notation is used to indicate that this new fermion will obtain a Dirac mass with the $\chi_f$ inside the $\overline{\mathbf{9}}$, as will be shown in section \ref{sec:Yukawa}. From this discussion, we can also see the advantage of using the $Sp(N)$ group instead of $SU(N)$, where the anomaly issue is more severe as discussed in Appendix~\ref{App:SU(3)}.

Next, for the $U(1)_5$ gauge symmetry to be anomaly-free, we also need to add new singlet fermions $\chi'_0$ that only carry $U(1)_5$ charge to each generation. Together with $\chi^c_f$, the anomalies can be properly canceled, with the charges of the new fermions under the $U(1)_5$ symmetry given by
\begin{align}
Q_5(\chi^c_f)=c/4~,\quad
Q_5(\chi'_0)=-c~,~\text{  where  }~
c=9^{1/3}~.
\end{align}

To sum up, we have an $SU(9)$ Dark GUT gauge group together with the $Sp(4)'\times U(1)_5$ gauge symmetry, forming the complete gauge sector of the model. The fermion content, including five different representations under the gauge group $SU(9)\times Sp(4)'\times U(1)_5$, denoted by $A,\Psi,\bar{F},\chi^c_f,\chi'_0$, is provided as follows:
\begin{align}
&3\times(\mathbf{36},\mathbf{1})_0=A_i=
\begin{pmatrix}
\chi_a \oplus \chi_0     &   \psi^c_c   &  \psi^c_w \\
-(\psi^c_c)^T   &   u^c      &  q      \\
-(\psi^c_w)^T   &   -q^T     &  e^c    \\
\end{pmatrix}_i
~\text{ , }~
3\times(\overline{\mathbf{9}},\mathbf{4})_{-1/4}=\Psi_i=
\begin{pmatrix}
\psi_s\oplus\psi_a\oplus\psi_0 \\ \psi_c \\ \psi_w 
\end{pmatrix}_i
~\text{ , }~\nonumber\\
&3\times(\overline{\mathbf{9}},\mathbf{1})_1=\bar{F}_i=
\begin{pmatrix}
\chi_f \\ d^c \\ \ell 
\end{pmatrix}_i
~\text{ , }~
3\times({\mathbf{1}},\mathbf{4})_{c/4}=\chi^c_{f,i}
~\text{ , }~
3\times({\mathbf{1}},\mathbf{1})_{-c}=\chi'_{0,i}~,
\label{AllF}
\end{align}
where $i=1,2,3$ for three generations of fermions. Among them, the $\chi_{a,i}$ and $\chi_{f,i}(\chi_{f,i}^c)$ will become the dark quarks in our dark QCD sector. An important feature, as a consequence of extending the SM fermion content, is that there are also three generations of dark quarks analogous to the SM quarks. This not only introduces a flavored dark sector \cite{Kile:2011mn,Batell:2011tc,Kamenik:2011nb,Agrawal:2011ze,Lopez-Honorez:2013wla,Kile:2013ola,Batell:2013zwa,Kile:2014jea,Renner:2018fhh}, but also allows new $CP$-violating sources that are relevant for cosmology. Further details regarding the masses and interactions of these fermions will be provided in the following subsections.


\subsection{Scalar content and symmetry breaking pattern}\label{sec:Scalar}

With the $SU(9)\times Sp(4)'\times U(1)_5$ gauge group constructed as described in the previous discussion, the next step is to break it down to the SM gauge group along with the dark color group. In this study, the symmetry breaking chain is given by
\begin{align}
SU(9) \times Sp(4)' \times U(1)_5~
\xrightarrow{~\Lambda_{\rm GUT}~} ~ 
&SU(2)_W \times U(1)_X \times SU(7)_{DU}\times Sp(4)'\times U(1)_5\\
\xrightarrow{~\Lambda_{\rm DU}~} ~
&SU(2)_W \times U(1)_Y \times SU(3)_C \times Sp(4)_D \times U(1)'_5~\\
\xrightarrow{~\Lambda_{\rm 5}~} ~
&SU(2)_W \times U(1)_Y \times SU(3)_C \times Sp(4)_D~\\
\xrightarrow{~\Lambda_{\rm EW}~} ~
&U(1)_{EM} \times SU(3)_C \times Sp(4)_D~.
\end{align}
Start with GUT breaking at the $\Lambda_{\rm GUT}$, where the $SU(2)_W$ gauge group is separated out. The $\Lambda_{\rm DU}$ represents the scale of Dark Unification (DU), where the QCD sector is unified with the dark QCD sector. The $\Lambda_{\rm 5}$ is the breaking scale of the $U(1)'_5$ gauge symmetry, which originates from the combination of $U(1)_5$ and a $U(1)$ subgroup of $SU(7)_{DU}$. The gauge group is ultimately broken down to the $SU(3)_C\times U(1)_{EM}$ along with the $Sp(4)_D$ dark color. Next, we examine each symmetry-breaking step, discussing the scalar sector required and the particle masses generated.

Start with the first breaking at the GUT scale $\Lambda_{\rm GUT}\sim 10^{16}$ GeV. A scalar field that transforms as an adjoint representation of $SU(9)$ (and as a singlet under $Sp(4)'\times U(1)_5$) is introduced with a non-zero vacuum expectation value (VEV) as
\begin{align}\label{GammaV}
\Gamma = (\mathbf{80},\mathbf{1})_0 ~\text{ with }~
\langle \Gamma \rangle = V\,T_{80}~,~\text{where}~
T_{80}=\frac{1}{6\sqrt{7}}
\begin{pmatrix}
2\,\mathbb{I}_{7\times 7}  &  0    \\
0  &  -7\,\mathbb{I}_{2\times 2}    \\
\end{pmatrix}~\text{of$~SU(9)\,$.}
\end{align} 
The VEV breaks the $SU(9)$ symmetry into $SU(7)_{DU}\times SU(2)_W\times U(1)_X$, where the generator of the $U(1)_X$ is simply $T_{80}$ shown above, i.e. $X=T_{80}$. A total of 28 gauge bosons become massive at this stage, including the leptoquarks, which are the same as those in the $SU(5)$ GUT, and lepto-dark-quarks, which couple dark quarks to leptons. We denote them as $V_{\rm LQ}$ and $V_{\rm LD}$. These gauge bosons, as well as the scalar bosons that are not eaten, all obtain masses from the $\langle \Gamma \rangle$ at the scale of $\Lambda_{\rm GUT}$.

The second step, which happens not too far from $\Lambda_{\rm GUT}$, is the breaking of $SU(7)_{DU}\times Sp(4)'\times U(1)_X\times U(1)_5$ around the scale $\Lambda_{\rm DU}\sim 10^{15}$ GeV. The breaking can be realized in two different ways. One can first break the $SU(7)_{DU}\times U(1)_X\times U(1)_5$ into the subgroup $Sp(4)\times SU(3)_C\times U(1)_Y\times U(1)'_5$ through a antisymmetric scalar field given by
\begin{align}
\Delta=(\mathbf{36},{\mathbf{1}})_{1/2}~\text{ with }~
\langle \Delta \rangle = \frac{f_\Delta}{\sqrt{2}}
\begin{pmatrix}
\mathbb{A}_{4\times 4}  &  0    \\
0  &  0    \\
\end{pmatrix}~,~\text{where}~
\mathbb{A}_{4\times 4}=
\begin{pmatrix}
0  &  \mathbb{I}_{2\times 2}    \\
-\mathbb{I}_{2\times 2}  &  0    \\
\end{pmatrix}.
\end{align} 
Once the breaking is realized, the quark-dark-quark field, denoted as $V_{\rm DQ}$, and the gauge bosons corresponding to the $SU(4)/Sp(4)$ coset, denoted as the dark coloron $G_D$, become massive. One of the $U(1)$ gauge symmetries is also broken, leaving two linear combinations of $U(1)_7\subset SU(7)_{DU}$ (with the generator $T_{48}$), $U(1)_X$, and $U(1)_5$ unbroken. One of them can be simply identified with the $U(1)_Y$ hypercharge, which can be expressed as
\begin{align}\label{T48}
Y=\sqrt{\frac{9}{7}}\,X-\sqrt{\frac{8}{21}}\,T_{48}~,~\text{where}~
T_{48} =\frac{1}{2\sqrt{42}}
\begin{pmatrix}
3\,\mathbb{I}_{4\times 4}  &  0    \\
0  &  -4\,\mathbb{I}_{3\times 3}    \\
\end{pmatrix} ~\text{of$~SU(7)_{DU}$ .}
\end{align}
The other unbroken $U(1)$ gauge symmetry, denoted as $U(1)'_5$, can be expressed as a linear combination of $U(1)_5$ and $U(1)_7$ with the charges following the relation
\begin{align}
Q'_5=Q_5+\sqrt{\frac{7}{6}}\,T_{48}~.
\end{align}
The charges of relevant fermions under the $U(1)'_5$ gauge symmetry are given by
\begin{align}\label{Q'5}
&Q'_5(\chi^a)=1/2~,&&Q'_5(\chi_f)=3/4~,&&Q'_5(\chi^c_f)=c/4~,&&Q'_5(\ell)=1~,\nonumber\\
&Q'_5(q)=-1/3~,&&Q'_5(u^c)=-2/3~,&&Q'_5(d^c)=4/3~,&&Q'_5(e^c)=0~,
\end{align}

Now with $Sp(4)\times SU(3)_C\times U(1)_Y\times U(1)'_5$ left from the DU group and $Sp(4)'$ remaining unaffected, we need to introduce another scalar field 
\begin{align}\label{Phi}
\Phi=(\mathbf{9},{\mathbf{4}})_{-1/4}~\text{ with }~
\langle \Phi \rangle = \frac{f_\Phi}{\sqrt{2}}
\begin{pmatrix}
\mathbb{I}_{4\times 4}  \\
0      \\
\end{pmatrix}~,
\end{align} 
which can further break the $Sp(4)\subset SU(7)_{DU}$ and $Sp(4)'$ down to the diagonal subgroup $Sp(4)_D$, our dark color group. This will lead to another set of massive dark coloron $G'_D$. The complete breaking at this scale can also be done by $\langle \Phi \rangle$ along. To simplify the symmetry breaking chain, we assume that the two scalars share the same VEV, $f_\Delta  = f_\Phi = f$. Therefore, all massive bosons mentioned in this part have masses of $\mathcal{O}(f)\sim 10^{15}$ GeV.

For the third step, we introduce a similar scalar $\Phi'$ but with different $U(1)_5$ charge
\begin{align}
\Phi'=(\mathbf{9},{\mathbf{4}})_{-1-c/4}~\text{ with }~
\langle \Phi' \rangle = 
\begin{pmatrix}
\langle \phi' \rangle  \\
0      \\
\end{pmatrix}
=\frac{f'}{\sqrt{2}}
\begin{pmatrix}
\mathbb{I}_{4\times 4}  \\
0      \\
\end{pmatrix},
\end{align} 
which is introduced to obtain the desired Yukawa coupling term and thereby generate the dark quark masses, as will be clarified in the next subsection. The VEV $\langle \Phi' \rangle$ shares a structure similar to $\langle \Phi \rangle$, but occurs at a much lower scale, $f'\sim 10^5$ GeV. Note that, under the previous symmetry breaking, $\Phi'$ is already decomposed into $(4,3,1)+(4,1,2)+(10,1,1)+(5,1,1)+(1,1,1)$ under the gauge group $Sp(4)_D\times SU(3)_C\times SU(2)_W$. To simplify the discussion, we assume that only the singlet complex scalar $\phi'=(1,1,1)$ remains around the scale $\Lambda_5$, while all other scalar fields acquire masses above $\Lambda_{\rm DU}$. The complex scalar singlet $\phi'$, however, carries a nonzero $U(1)'_5$ charge $c_{f'}\equiv Q'_5(f')=(-3-c)/4\approx 1.3$. Consequently, when $\phi'$ acquires a VEV $f'\sim 10^5$ GeV, it breaks the $U(1)'_5$ gauge symmetry, resulting in a massive $Z'$ boson with a mass given by $M_{Z'}=c_{f'}g_{Z'}f'\sim \mathcal{O}(10)$ TeV, which is phenomenologically relevant. In addition to the axial mode absorbed by the $Z'$ boson, the radial mode, which couples to the dark quarks, will also play a role in the two-loop beta function, as will be mentioned in Appendix~\ref{App:Twoloop}.

Lastly, we also need to generate the electroweak symmetry breaking as well as the observed SM fermion masses. This requires the inclusion of Higgs doublets. In our Dark GUT model, two Higgs doublets are required, originating from the multiplets given by
\begin{align}
H_d\subset H_{9}=(\mathbf{9},\mathbf{1})_{1}~,\quad
H_u\subset H_{126}=(\mathbf{126},\mathbf{1})_0~.
\end{align}
This leads to a standard Type-II 2HDM \cite{Branco:2011iw}, which is a direct consequence of extending the $SU(5)$ GUT. To simplify the discussion, we again assume only the two Higgs doublets exist at the electroweak scale, while other extra scalars have masses above $\Lambda_{\rm DU}$.

To sum up, the total scalar content is given as follows (according to different purposes):
\begin{itemize}
\item Realizing the GUT symmetry breaking: $\Gamma = (\mathbf{80},\mathbf{1})_0$~,
\item Realizing the DU symmetry breaking: $\Delta=(\mathbf{36},{\mathbf{1}})_{1/2}$~, $\Phi=(\mathbf{9},{\mathbf{4}})_{-1/4}$~,
\item Realizing the $U(1)'_5$ symmetry breaking: $\Phi'=(\mathbf{9},{\mathbf{4}})_{-1-c/4}$~,
\item Realizing the EWSB and SM fermion masses : $H_d\subset(\mathbf{9},\mathbf{1})_1$~, $H_u\subset(\mathbf{126},\mathbf{1})_0$~.
\end{itemize}
Besides the scale-setting scalars above, we also add three additional scalars, $\Gamma_i = (\mathbf{80}, \mathbf{1})_0$, around the scale $\Lambda_5$. Their only purpose is to help unify the running of the $U(1)_Y$ coupling constant, as will be discussed further in section \ref{sec:Spectrum}. We emphasize that these fields play no role in other mechanisms and can be replaced by other fields which have similar quantum number and can serve the same purpose, such as gauginos in supersymmetric theories.

\subsection{Yukawa interaction terms and fermion masses}\label{sec:Yukawa}

With the complete fermion and scalar contents given, we can then discuss the Yukawa interactions among them. Obeying the $SU(9) \times SU(4)' \times U(1)_5$ gauge symmetry, the allowed Yukawa interaction terms are given as 
\begin{align}\label{Yukawa}
-{\cal L}_{\rm Yukawa} = 
Y_\Delta \Psi\Psi\Delta + Y A \Psi \Phi^* + 
Y' \bar{F} \chi^c_f \Phi'  + Y_u A A H_{126}
+ Y_d A \bar{F} H_{9}^* +\text{h.c.}~,
\end{align}
where we suppress the flavor indices here for simplicity. After the scalar fields acquire their VEVs, as discussed in the previous section, the Yukawa terms then become mass terms for the corresponding fermions.

Starting from the first term, we have
\begin{align}
-{\cal L}_{\Delta} = 
Y_{\Delta,i,j} \langle \Delta \rangle \left(\psi_{s,i}\psi_{s,j}+\psi_{a,i}\psi_{a,j}+\psi_{0,i}\psi_{0,j}\right)+\text{h.c.}~,
\end{align}
where $Y_{\Delta,i,j}$ is a $3\times 3$ matrix. The coupling to the vacuum $\langle \Delta \rangle$ gives the three species Majorana masses written as
\begin{align}\label{Majmass}
M_\Delta = Y_{\Delta} \langle \Delta \rangle ~\to~
M_\Delta = U_\Delta^*~ {M_\Delta^{\text{diag}}}~ U_\Delta^\dagger= 
U_\Delta^*~
\begin{pmatrix}
0.3\,f   &  0  &  0   \\
0  &  0.3\,f  &  0  \\
0   &  0 &  0.3\,f   \\
\end{pmatrix}
U_\Delta^\dagger~,
\end{align}
where we have the mass matrix $M_\Delta$ diagonalized through a unitary matrix $U_\Delta$. For simplicity, we assume degenerate mass eigenvalues for the three generations, which gives all $\psi_{s,i}$, $\psi_{a,i}$, and $\psi_{0,i}$ the same mass, $M_s = M_a = M_0 = 0.3 \times 10^{15}$ GeV. The factor $0.3$ is introduced to achieve the desired mass hierarchy, ensuring that the fermions at the DU scale are lighter than the bosons at the DU scale. This distinction is relevant for cosmology, as will be discussed in section \ref{sec:COS}.

Next, for the second term, the $\Phi$ field couples $\chi\subset A$ and $\psi\subset \Psi$ through the terms
\begin{align}\label{LPhi}
-{\cal L}_{\Phi} = 
Y_{i,j} \langle \Phi \rangle \left(\psi^c_{c,i}\psi_{c,j}+\psi^c_{w,i}\psi_{w,j}+\chi_{a,i}\psi_{a,j}+\chi_{0,i}\psi_{0,j}\right)+\text{h.c.}~,
\end{align}
which gives Dirac masses $M_\psi = Y \langle \Phi \rangle$ to the exotic SM charged fermions, as we previously mentioned. Now, we need two unitary matrices, $U_{\psi}$ and $V_{\psi}$, to diagonalize $M_\psi$, as
\begin{align}
M_\psi = U_{\psi}~ {M_\psi^{\text{diag}}}~ V_{\psi}^\dagger 
\implies
M_\psi^{\text{diag}}=
\begin{pmatrix}
10^{7}   &  0  &  0   \\
0  &  10^{11}  &  0  \\
0   &  0 &  10^{15}   \\
\end{pmatrix}\text{ GeV.}
\label{Ymatrix}
\end{align}
Here, we impose a nontrivial flavor structure, different from the Majorana masses $M_\Delta$. Analogous to the SM, we expect hierarchical masses among three generations as 
\begin{align}
m_{c,3}=m_{w,3}\sim 10^{15} \text{ GeV}~,~~
m_{c,2}=m_{w,2}\sim 10^{11} \text{ GeV}~,~~
m_{c,1}=m_{w,1}\sim 10^{7} \text{ GeV}~,
\end{align}
where $m_c$ and $m_w$ stands for the masses of Dirac fermions $\psi_{c}$ and $\psi_{w}$. In this work, we assume a similar flavor pattern to the SM ($m_3 \gg m_2 \gg m_1$). However, the flavor structure could also be significantly different, or even inverted ($m_1 \gg m_2 \gg m_3$), compared to that of the SM, which would lead to different expectations for the phenomenology connecting the two sectors. We leave such considerations for future study.

For the remaining two mass terms in eq. \eqref{LPhi}, $\chi_{a}\psi_{a}$ and $\chi_{0}\psi_{0}$, since $\psi_a$ and $\psi_0$ already acquire $\mathcal{O}(10^{15})$ GeV Majorana masses, the $\chi_a$ and $\chi_0$ fields will obtain masses through the seesaw mechanism, resulting in a quadratic mass hierarchy\footnote{We assume that the unitary matrices are similar in both sectors to avoid a seesaw enhancement \cite{Smirnov:1993af,Akhmedov:2003dg} and to achieve the desired quadratic mass hierarchy. Otherwise, for example, if $M_\Delta$ is anti-diagonal in the basis where $M_\psi$ is diagonal, one would instead obtain a reduced mass hierarchy, with nearly degenerate masses for the light fermions \cite{Chung:2023rie}.} as
\begin{align}
m_{a,3}=m_{0,3}\sim 10^{15} \text{ GeV}~,\quad
m_{a,2}=m_{0,2}\sim 10^{7} \text{ GeV}~,\quad
m_{a,1}=m_{0,1}\sim 0.1 \text{ GeV}~,
\end{align}
where $m_a$ and $m_0$ are the Majorana masses of $\chi_a$ and $\chi_0$. The lightest antisymmetric fermion $\chi_{a,1}$ is intentionally arranged to have a mass below the GeV scale, such that it can serve as a light dark quark, leading to a richer dark QCD sector.


The third term in eq. \eqref{Yukawa} gives masses to the other dark quarks $\chi_f$ and $\chi^c_f$ as
\begin{align}
-{\cal L}_{\Phi'} = 
Y'_{i,j} \langle \Phi' \rangle \chi_{f,i}\chi^c_{f,j}+\text{h.c.}~.
\end{align}
Again, for the mass matrix $M_f = Y' f'$ with $f'\sim 10^5$ GeV, we expect a hierarchical mass spectrum, albeit milder than that of $M_\psi$, given by
\begin{align}
M_f = U_{f}~ {M_f^{\text{diag}}}~ V_{f}^\dagger 
\implies
M_f^{\text{diag}}=
\begin{pmatrix}
m_{f,1}   &  0  &  0   \\
0  &  m_{f,2}  &  0  \\
0   &  0 &  m_{f,3}   \\
\end{pmatrix}=
\begin{pmatrix}
0.1   &  0  &  0   \\
0  &  10^{2}  &  0  \\
0   &  0 &  10^{5}   \\
\end{pmatrix}\text{ GeV}~,
\label{Ymatrix}
\end{align}
where $m_{f,i}$ are the masses of Dirac fermions $\chi_{f,i}$. Unlike the $\chi_{a,i}$ discussed previously, the $\chi_{f,i}$ acquire their masses from the scale $\Lambda_5 \sim 10^5$ GeV, allowing them to naturally spread down to the GeV scale. Another noteworthy difference is that the $\chi_a$ are Majorana fermions, whereas the $\chi_{f}$ are Dirac fermions. This distinction originates from the fact that the antisymmetric fermions $\chi_a$ belong to real representations, while the fundamental fermions $\chi_{f}$ belong to pseudo-real representations.

The last two terms in eq. \eqref{Yukawa} give
\begin{align}\label{Higgs_int}
-{\cal L}_{H} = Y_u\, q u^c H_u + Y_d\, q d^c H_d + Y_d\, \ell e^c H_d + Y_d\, \psi_w^c \chi_f H_d  +\text{h.c.}~,
\end{align}
which include the SM Yukawa couplings in the form of a Type-II 2HDM with identical couplings for down-type quarks and charged leptons, as predicted by the $SU(5)$ GUT. In this work, we do not address the flavor problem in $SU(5)$ GUT, but assume that the SM fermion masses can be correctly reproduced at low energies by some mechanism, while noting that its resolution does not affect the physics of our interest. Besides the first three terms required from the SM, there is one last term which introduces a small mixing of $\mathcal{O}(v/m_w)$ between the electrically neutral component $\psi_w^0$ of heavy exotic fermions $\psi_w$ and dark quarks $\chi_f$.

\subsection{Mass spectrum, running couplings, and confinement scales}\label{sec:Spectrum}

\begin{table}[t]
\centering
\begin{tabular}{c|c|c|c|c|c|}
GeV		& Scale				& Scalar		& Gauge boson	& Exotic fermion	& Dark quark  \\ \hline
$10^{16}$	& $\Lambda_{\rm GUT}$ 	& $\Gamma$	& $V_{\rm LQ},\,V_{\rm LD}$	&  &  \\ 
$10^{15}$	& $\Lambda_{\rm DU}$		& $\Delta,\,\Phi$	& $V_{\rm DQ},\,Z'_{X},\,G^{(\prime)}_D$	& $\psi_{c,3},\,\psi_{w,3},\,\psi_{s,i},\,\psi_{a,i}$	& $\chi_{a,3}~\qquad$   \\
&&&&&\\
$10^{11}$	& 	& & & $\psi_{c,2},\,\psi_{w,2}$	&    \\
&&&&&\\
$10^{7}$	& 	& & & $\psi_{c,1},\,\psi_{w,1}$	& $\chi_{a,2}~\qquad$   \\
\rule{0pt}{15pt}$10^{5}$	& $\Lambda_{\rm 5}$		& $\phi',\,\Gamma_i$		& $Z'$	& 	& $\qquad~\chi_{f,3}$ \\
 &&&&&\\
\rule{0pt}{15pt}$10^{2}$	& $\Lambda_{\rm EW}$		& $H_u,\,H_d$	& $W,\,Z$	& 	&  $\qquad~\chi_{f,2}$  \\[5pt]
$1$		& $\Lambda_{\rm DC}\sim \Lambda_{\rm QCD}$		& & & &   \\
$0.1$		& 	& & & & $\chi_{a,1},~\chi_{f,1}$  \\
\end{tabular}
\caption{The overall spectrum of relevant scales and new particles. See the text for more details. \label{tab:spectrum}}
\end{table}

With all the mass terms given, we present the complete mass spectrum in table~\ref{tab:spectrum}. The spectrum features six relevant scales, including $\Lambda_{\rm GUT}$, $\Lambda_{\rm DU}$, $\Lambda_{\rm 5}$, $\Lambda_{\rm EW}$, $\Lambda_{\rm DC}$, and $\Lambda_{\rm QCD}$, as shown in the second column, and their corresponding values in the first column are given in units of GeV. The first four scales all feature the corresponding scalar fields that acquire a nonzero VEV, listed in the third column. Each VEV breaks a part of the gauge symmetry, introducing massive gauge bosons at the corresponding scale, as indicated in the fourth column. Together, they break the gauge symmetry all the way from the Dark GUT group $SU(9) \times Sp(4)' \times U(1)_5$ to the $Sp(4)_D\times SU(3)_C\times U(1)_{EM}$.

The last two columns present new fermions charged under $Sp(4)_D$. The exotic fermions obtain their masses from the scale $\Lambda_{\rm DU}$. The spectrum of $\psi_{c,i}$ and $\psi_{w,i}$ spreads out over orders of magnitude due to hierarchical Yukawa couplings, analogous to the SM fermions. The dark quarks, $\chi_{a,i}$, despite obtaining masses from $\Lambda_{\rm DU}$, receive further suppression due to the seesaw mechanism, leading to the quadratic mass hierarchy with the lightest $\chi_{a,1}$ around $0.1$ GeV. The other dark quarks, $\chi_{f,i}$, obtain their Dirac masses from the scale $\Lambda_{\rm 5}$, which is much lower, and also spread down to the $0.1$ GeV range through the hierarchical Yukawa couplings. Other new fermions, including $\psi_{0,i}$, $\chi_{0,i}$, and $\chi'_{0,i}$, are singlets under the unbroken gauge groups and play no role in our discussion. We will simply ignore them in the following discussion.

Next, we look into the details of each scale. The values of these scales are determined by considering experimental constraints, phenomenological requirements, and the running of coupling constants. Starting from the GUT scale, which features leptoquark fields $V_{\rm LQ}$ that mediate proton decay, we set $\Lambda_{\rm GUT} = V = 10^{16}$ GeV to satisfy current limits from Super-Kamiokande \cite{Super-Kamiokande:2020wjk}. Additionally, the lepto-dark-quark fields $V_{\rm LD}$, which couple to the lepton and the dark matter, also acquire their masses at the GUT scale.

\begin{figure}[tbp]
\centering
\includegraphics[width=0.9\textwidth]{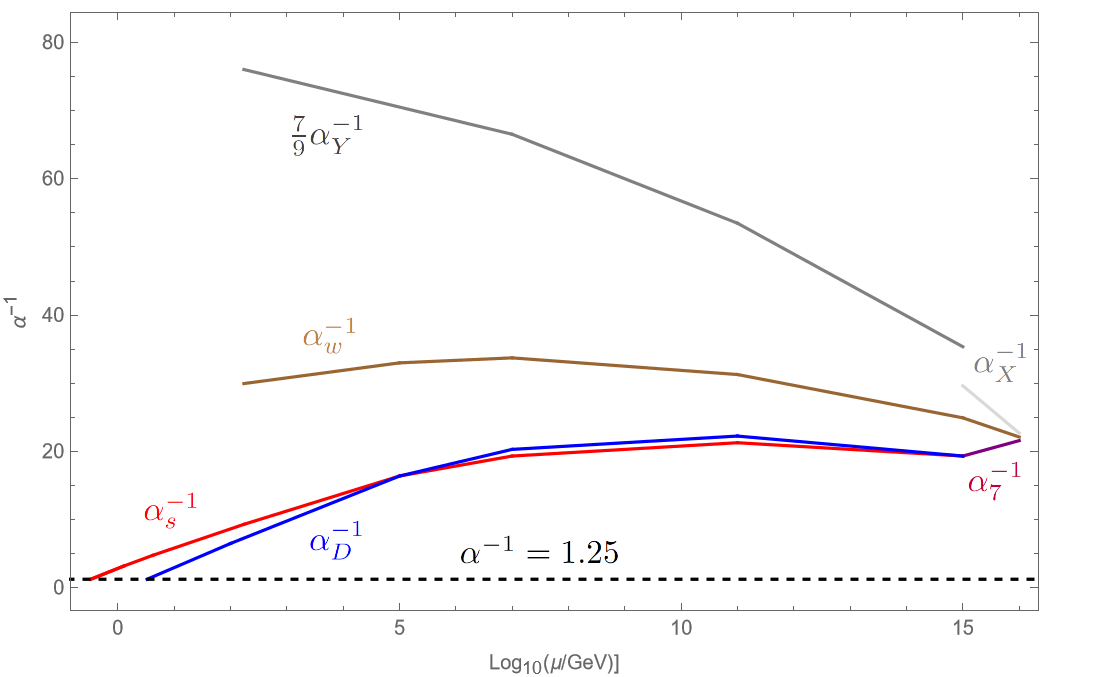}
\caption{\label{fig:running} One-loop running of the coupling constants originated from the $SU(9)$ gauge group, including $\alpha_s$, $\alpha_w$, $\alpha_Y$ for the three SM gauge groups and $\alpha_D$ for the dark color group. The running coupling constants of unified group $SU(7)_{\rm DU}\times U(1)_X$ above the $\Lambda_{\rm DU}$ scale are also shown, which ultimately unify at the $\Lambda_{\rm GUT}$ scale. In this figure, the $Sp(4)'$ and $U(1)'_5$ couplings are not included since they are not part of $SU(9)$ gauge group. Furthermore, we assume the $Sp(4)'$ coupling to be infinite in this calculation. The impact of this coupling, especially on dark color confinement scale, will be discussed in the end of the subsection \ref{sec:Spectrum}.}
\end{figure}

The second scale is the scale of dark unification, above which the QCD and dark QCD are unified into the dark unification group $SU(7)_{DU}$. Due to the strong running of $SU(7)_{DU}$ gauge group, the splitting between $\Lambda_{\rm GUT}$ and $\Lambda_{\rm DU}$ should be small to match the coupling difference between the strong coupling $\alpha_s$ and the weak coupling $\alpha_w$ at low energies. We also assume that the scalars charged under the weak interaction are heavier in order to split the two scales further. The result of the running couplings is summarized in figure \ref{fig:running}, where we get $\Lambda_{\rm DU}=f= 10^{15}$ GeV, one order of magnitude below $\Lambda_{\rm GUT}=10^{16}$ GeV.

The scale $\Lambda_{\rm DU}$ is also the most important scale in this framework, as most fermions acquire their masses from it. The $\psi_a$, which plays an important role in cosmology, receive Majorana masses $M_a$ at this scale. On the other hand, for $\psi_c$ and $\psi_w$, their masses spread all the way from $10^{15}$ GeV to $10^{7}$ GeV due to their hierarchical Yukawa couplings. The dark quarks, $\chi_a$, which obtain masses through the seesaw mechanism, have a quadratically hierarchical spectrum, leading to the lightest $\chi_{a}$ with a mass $\sim 0.1$ GeV even though its mass originates from $f= 10^{15}$ GeV. Besides fermions, the scale is also responsible for the masses of many bosons. In addition to $\Delta$ and $\Phi$, which acquire their VEVs from the scale, we also expect other scalar components of $\Phi'$, $H_{9}$, and $H_{126}$ to reside around this scale.\footnote{In this work, we do not include the doublet-triplet splitting problem or the potential hierarchy problem to simplify the discussion. We assume these problems can be solved without affecting our mechanism.} There are also many massive gauge bosons. Among them, the $V_{\rm DQ}$ gauge boson, which mediates transitions between dark-colored fermions and SM quarks, plays an important role in the decay of exotic fermions and thus in cosmology. For a viable scenario for cosmology, even though many particles acquire masses of $\mathcal{O}(10^{15})$ GeV, we must further impose a hierarchy among them to determine their decay properties. In this study, we assume
\begin{align}
{\rm Scalar}~\Delta,\,\Phi~>~
{\rm Vector}~V_{\rm DQ},\,Z'_{X},\,G^{(\prime)}_D~>~
{\rm Fermion}
~\psi_{a,i},\,\psi_{s,i},\,\psi_{c,3},\,\psi_{w,3},\,\chi_{a,3}~.
\end{align}
As the gauge boson masses are determined by the gauge coupling $M_{\rm DQ}=\frac{1}{2}g_7f\sim 0.4\,f$. We therefore choose a smaller Yukawa coupling in eq.~\eqref{Majmass} and the fermion masses $M_a=0.3\,f$ as the benchmark value.

Moving down to the scale $\Lambda_{\rm 5}$ with $f' = 10^{5}$ GeV, this value is chosen for phenomenological and cosmological purposes. One is achieving the desired dark quark spectrum of $\chi_f$, with the lightest $\chi_{f,1}$ below the GeV scale as will be discussed in section \ref{sec:DS}. The other purpose is meeting the requirement for the annihilation of symmetric abundance, which is an important step in cosmology. In our model, this process is mediated through the $Z'$ boson, whose effect is determined by the $f'$. More detail will be discussed in section~\ref{sec:COS}. Both of these requirements necessitate an intermediate scale not too far from the electroweak scale. Therefore, we set it near the acceptable upper bound at $10^{5}$ GeV.

Besides, we also add three adjoint scalars $\Gamma_i = \mathbf{80}$ under $SU(9)$ around this scale, where we expect only the components $\Gamma_d = (\mathbf{10}, \mathbf{1}, \mathbf{1})$, $\Gamma_c = (\mathbf{1}, \mathbf{8}, \mathbf{1})$, and $\Gamma_w = (\mathbf{1}, \mathbf{1}, \mathbf{3})$ under $Sp(4)_D \times SU(3)_C \times SU(2)_W$ to have masses around $10^5$ GeV, while other components are much heavier. The only purpose of these fields is to slow down the running of gauge couplings other than $U(1)_Y$, playing a role similar to that of gauginos in supersymmetric GUT. Although they are rather light, their lack of direct Yukawa couplings with SM fermions prevents them from introducing dangerous observables. These redundant fields play no role in other mechanisms and can therefore be waived by working on the supersymmetric version of Dark GUT, which we leave for future study.



\begin{table}[]
\centering
\begin{tabular}{|c|c|c|c|c|c|c|}
\hline
coefficient  & pure gauge & [$1$, $m_t$] & [$m_t$, $10^5$] & [$10^5$, $10^7$]  & [$10^7$, $10^{11}$] & [$10^{11}$, $10^{15}$]  \\ \hline
$b_{s}$   & $11$  & $9-7\sfrac{2}{3}$      & $7$  & $4$  & $1\sfrac{1}{3}$ & $-1\sfrac{1}{3}$ \\ \hline
$b_{D}$    & $11$  & $9\sfrac{2}{3}$   & $9$  & $5\sfrac{1}{3}$ & $1\sfrac{1}{3}$ & $-2$ \\ \hline
\end{tabular}
\caption{The comparison of the one-loop beta function coefficients between QCD and DC across different intervals of scales in units of GeV.\label{tab:beta}}
\end{table}

The last two scales, $\Lambda_{\rm DC}$ and $\Lambda_{\rm QCD}$, are the confinement scales of corresponding gauge groups. Below $\Lambda_{\rm DU}$, the couplings of dark color and color groups separate and change independently at one-loop level.\footnote{At two-loop level, the two running couplings become correlated due to the presence of bifundamental fermions $\psi_c$. The complete calculation is shown in Appendix~\ref{App:Twoloop}. There are also bifundamental scalars in the model but we have assumed them to be comparable or heavier than $\Lambda_{\rm DU}$ so they do not affect the coupling running below the $\Lambda_{\rm DU}$ scale.} The trajectories of the two running couplings considering one-loop beta functions are shown in figure \ref{fig:running}. Taking $\alpha_{s}^{-1}(\Lambda_{\rm QCD} = 0.33\ \text{GeV}) \sim 1.25$ as a reference for estimating the confinement scale, we find that, for dark color, $\alpha_{D}^{-1}(\Lambda_{\rm DC}) \sim 1.25$ when $\Lambda_{\rm DC} \sim 3.3\ \text{GeV}$, which is ten times greater than $\Lambda_{\rm QCD}$. It is important to emphasize that this ratio is already a great success considering that the separated running starts at $\Lambda_{\rm DU} \sim 10^{15}\ \text{GeV}$. To understand this, we can examine the coefficient of the one-loop beta function in detail, as shown in table \ref{tab:beta}. We find that the two one-loop beta-function coefficients, $b_{s}$ and $b_{D}$, are comparable across the entire energy regime and even identical over one of the intervals.

Such a similarity is not a mere coincidence but arises from two factors. The first is the identical $C_2(G)$ factor shared by $SU(3)_C$ and $Sp(4)_D$, which we imposed by hand in section~\ref{sec:Sp4}. The other, however, comes from the similar fermion content,\footnote{Notice that the scalar fields $\Gamma_d$ and $\Gamma_c$ that appear at the $10^5$ GeV give the same contribution $\Delta b=-3$ to both sectors. Therefore, they do not play any role in this comparison.} which is the direct consequence of dark grand unification. We would also like to emphasize that, in the limit where all the $\psi$ fermion fields decouple and all the dark quarks are taken into the massless limit, the one-loop beta-function coefficient of dark color is given by
\begin{align}\label{bDC}
b_{D}=\frac{11}{3}\times 3-\frac{2}{3}\times 3 \times (1\times 1 +2\times\frac{1}{2})=11-4=7~.
\end{align}
It is exactly the same as the QCD running in the energy regime above the top quark mass, which stems from the fact that the dark quarks $\{\chi_a,\chi_f,\chi_f^c\}_i$ have the same contribution to the $Sp(4)_D$ running as the SM quarks $\{q,u^c,d^c\}_i$ do to the $SU(3)_C$ running.\footnote{The beta-functions are different at two-loop level as will be discussed in Appendix~\ref{App:Twoloop}. However, as the deviation is suppressed by a loop factor, the conclusion still hold.}


\begin{figure}[tbp]
\centering
\includegraphics[width=0.83\textwidth]{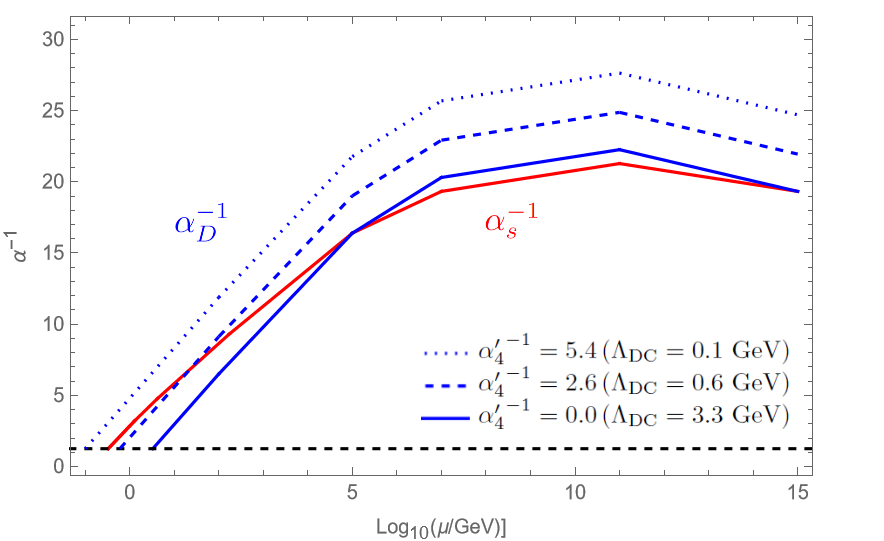}
\caption{\label{fig:alpha4} One-loop running of ${\alpha_s}$ and ${\alpha_D}$ according to different ${\alpha'_4}$.}
\end{figure}

Yet, we have not taken the effect of $Sp(4)'$ into account. Since the actual $Sp(4)_D$ is the diagonal subgroup of $Sp(4)\subset SU(7)_{DU}$ and $Sp(4)'$, the actual coupling of $Sp(4)_D$ should be $\alpha_{D}^{-1}={\alpha'_{4}}^{-1} + \alpha_{7}^{-1}$. In figure \ref{fig:running}, we directly use $\alpha_{D}=\alpha_{7}$ at the $\Lambda_{\rm DU}$ scale, which is equivalent as taking the coupling $\alpha'_{4}$ of $Sp(4)'$ to infinity. Once we consider a finite $\alpha'_{4}$, the whole curve of $\alpha_{D}^{-1}$ running shifts upwards, resulting in a rather small $\Lambda_{\rm DC}$, making it even closer to $\Lambda_{\rm QCD}$, as shown in figure \ref{fig:alpha4}. However, if ${\alpha'_{4}}$ is too weak, the curve will shift too much, producing a $\Lambda_{\rm DC}$ that is too small. Moreover, if ${\alpha'_{4}}^{-1}$ is much greater than $\alpha_{7}^{-1}$, the comparable initial value between $\alpha_s$ and $\alpha_D$ at the $\Lambda_{\rm DU}$ scale will no longer be true, which will also ruin the idea of applying unification. Therefore, to preserve the benefit of unification, a strongly coupled $Sp(4)'$ is required in this theory. An attractive possibility is that $Sp(4)'$ becomes strongly coupled and confines at the scale $\Lambda_{\rm DU}$. If the charged fermions form a "color-flavor locked" condensate \cite{Raby:1979my}, which behaves like the scalar $\Phi$ in eq.~\eqref{Phi}, it can then trigger dynamical symmetry breaking. However, a chiral strongly coupled theory is required and the existence of the phases remain under debate  \cite{Appelquist:2000qg,Csaki:2021xhi,Bolognesi:2021jzs,Smith:2021vbf,Karasik:2022gve,Li:2025tvu}, placing this possibility beyond the scope of the present study.

To sum up, the comparable DC and QCD confinement scales and thus comparable dark matter and baryon masses is possible under the framework of dark unification with few more requirements, including
\renewcommand{\labelenumi}{(\arabic{enumi})} 

\begin{enumerate}

    \item The choice of the dark color group $Sp(4)_D$ with the same quadratic Casimir as $SU(3)_C$
    \item The dark quarks $\{\chi_a, \chi_f, \chi_f^c\}_i$ which give the same contribution to the $Sp(4)_D$ running as the SM quarks $\{q, u^c, d^c\}_i$ give to the $SU(3)_C$ running
    \item A strongly coupled $Sp(4)'$ gauge group to ensure a similar initial condition at $\Lambda_{\rm DU}$
    
\end{enumerate}
Some features arise as direct consequences of unification, such as similar initial conditions and identical fermion family numbers. However, certain ad hoc assumptions are still required to achieve comparable confinement scales, reflecting the nontrivial challenge of aligning the running coupling constants of two different gauge groups.


\subsection{Conserved $U(1)$ symmetries}\label{sec:U(1)}

To identify the conserved $U(1)$ global symmetries, we start with the Lagrangian at a low energy, which contains only the SM and dark quarks $\{\chi_{a},\chi_{f},\chi^c_{f}\}$ as
\begin{align}\label{LlowE}
{\cal L}_{\rm LE} = {\cal L}_{\rm SM} - \frac{1}{4}G_{\mu\nu}^2+i\bar{\chi}_a\slashed{D}{\chi}_a-m_a{\chi}_a^c{\chi}_a+i\bar{\chi}_f\slashed{D}{\chi}_f
-m_f \bar{\chi}_{f}\chi^c_{f}+\text{h.c.}~.
\end{align}
Besides the well-known \textbf{baryon number $B$}  and \textbf{lepton number $L$} within ${\cal L}_{\rm SM}$, which are accidental $U(1)$ symmetries up to non-perturbative effects. In the dark sector, we can also define a new \textbf{dark baryon number $D$}, analogous to baryon number $B$, where 
\begin{align}
Q_D({\chi}_f)=1/2~,\quad Q_D({\chi}_f^c)=-1/2~,\quad Q_D({\chi}_a)=Q_D({\rm SM})=0~.
\end{align}
Notice that, for the dark quark $\chi_a$, due to its Majorana mass, no $U(1)$ charge can be assigned. For $\chi_f$ and $\chi_f^c$, they have opposite charges, forming a Dirac fermion. The charge is normalized to $1/2$ to match the correct dark baryon number, which will be discussed in the next section. We can easily check that the dark baryon number $D$ is also a global $U(1)$ symmetry in the low-energy Lagrangian \eqref{LlowE}.

However, at high energies, with new bosons and interactions from the grand unified theories, all of these symmetries are broken in some way. Nevertheless, in the $SU(5)$ GUT, the $B-L$ is conserved, which can be treated as a linear combination of two $U(1)$'s as
\begin{align}
B-L=\frac{1}{5}\left(4\,Y+Q\right)~.
\end{align}
The $Y$ is simply the hypercharge, which is gauged, while $Q$ is the charge associated with the $U(1)$ global symmetry of the $SU(5)$ GUT, where $Q(\mathbf{10}) = 1$ and $Q(\overline{\mathbf{5}}) = -3$. The two $U(1)$'s are separately conserved after the GUT breaking but are broken to the $B-L$ after the SM Higgs doublet acquires a VEV at the electroweak scale. In the $SU(9)$ Dark GUT, we aim to identify a similar conserved $U(1)$ symmetry, which is expected to encompass not only the baryon number $B$ and lepton number $L$, but also the dark baryon number $D$.

\begin{table}
\centering
\begin{tabular}{|c|c|c|c|c|c|c|c|c|c|c|c|}\hline
 & \multicolumn{5}{c|}{Fermion}
 & \multicolumn{6}{c|}{Scalar}  \\ \hline
Symmetry & $A$ & $\bar{F}$ & $\Psi$ & $\chi^c_f$ & $\chi'_0$ & $\Gamma$ & $\Delta$ & $\Phi$ & $\Phi'$ & $H_{9}$ & $H_{126}$  \\ \hline
$Q$  & $5$ & $-7$ & $-7$ & $9$ & $0$ & $0$ & $14$ & $-2$ & $-2$ & $-2$ & $-10$\\ \hline
$Q'$ & $0$ & $-4$ & $1$ & $3$ & $0$ & $0$ & $-2$ & $1$ & $1$ & $-4$ & $0$\\ \hline
\end{tabular}
\caption{\label{tab:U(1)G} The charges of fermions and scalars under the global $U(1)$ symmetries of the Dark GUT.}
\end{table}

We first return to the $SU(9)$ chiral theory with antisymmetric and anti-fundamental representations. The anomaly-free fermion content features $SU(5)_G \times U(1)_G$ global symmetry. The $Q$ is now associated with the $U(1)_G$ global symmetry of the $SU(9)$ Dark GUT, where $Q(\mathbf{36}) = 5$ and $Q(\overline{\mathbf{9}}) = -7$. The charges for the complete matter content are shown in table \ref{tab:U(1)G}. Furthermore, we also identify another global $U(1)'_G$ symmetry, which is a subgroup of $SU(5)_G$, with charge $Q'$ shown in table \ref{tab:U(1)G}. Both $U(1)$ global symmetries are preserved under the $SU(9)$ gauge interactions and Yukawa interactions in eq. \eqref{Yukawa}. However, they are both broken by the scalar VEVs.

To find the conserved $U(1)$ symmetry under all the scalar VEVs, we need to further include the $U(1)$ subgroups of $SU(9)$. One relevant subgroup is $U(1)'_9$ with a generator 
\begin{align}
T'_{80}=
\begin{pmatrix}
5\,\mathbb{I}_{4\times 4}  &  0    \\
0  &  -4\,\mathbb{I}_{5\times 5}    \\
\end{pmatrix},
\end{align} 
under which the dark-colored fermions and the SM fermions have different charges. We use $T'_{80}$ to distinguish from the $T_{80}$ in eq.~\eqref{GammaV} which aligns with the VEV at the GUT scale. One can check that it is a linear combination of $T_{80}$ and $T_{48}$ in eq.~\eqref{T48}.

\begin{table}
\centering
\begin{tabular} {|c|c|c|c|c|c|c|c|c|c|c|c|c|c|c|}\hline
& \multicolumn{6}{c|}{$A=(\mathbf{36},\mathbf{1})_0$}
& \multicolumn{3}{c|}{$\bar{F}=(\overline{\mathbf{9}},\mathbf{1})_1$}
& \multicolumn{3}{c|}{${\Psi}=(\overline{\mathbf{9}},\mathbf{4})_{-1/4}$} & \\ \hline
Symmetry & $\chi_a$ & $\psi^c_c$ & $\psi^c_w$ & $u^c$ & $q$ & $e^c$ & $\chi_f$ & $d^c$ & $\ell$ & $\psi_a$ & $\psi_c$ & $\psi_w$ & $\chi^c_f$  \\ \hline
${D}$  & $0$ & $-\frac{1}{2}$ & $-\frac{1}{2}$ & $0$ & $0$ & $0$ & $\frac{1}{2}$ & $0$ & $0$ & $0$ & $\frac{1}{2}$ & $\frac{1}{2}$ & $-\frac{1}{2}$ \\ \hline
${B}$  & $0$ & $-\frac{2}{3}$ & $0$ & $-\frac{1}{3}$ & $\frac{1}{3}$ & $0$ & $0$ & $-\frac{1}{3}$ & $0$ & $0$ & $\frac{2}{3}$ & $0$ & $0$ \\ \hline
${L}$  & $0$ & $0$ & $0$ & $0$ & $0$ & $-1$ & $0$ & $0$ & $1$ & $0$ & $0$ & $0$ & $0$ \\ \hline
${D-(B-L)}$  & $0$ & $\frac{1}{6}$ & $-\frac{1}{2}$ & $\frac{1}{3}$ & $-\frac{1}{3}$ & $-1$ & $\frac{1}{2}$ & $\frac{1}{3}$ & $1$ & $0$ & $-\frac{1}{6}$ & $\frac{1}{2}$ & $-\frac{1}{2}$ \\ \hline
\end{tabular}
\caption{\label{tab:U(1)} The charge of fermions under $U(1)$ global symmetries, $D$, $B$, $L$, and $D-(B-L)$.}
\end{table}

With all the ingredients, we find a new $U(1)$ symmetry with the charge given by
\begin{align}
\tilde{Q}=\frac{1}{18}\left(2\,Q+9\,Q'-T'_{80}\right)~
\end{align}
is conserved under all the scalar VEVs except the SM Higgs.\footnote{Another linear combination with the charge given by
\begin{align}
X=-\frac{1}{45}\left(8\,Q-9\,Q'+5T'_{80}\right)
\end{align}
is studied in \cite{Barr:2011cz,Barr:2013tea,Barr:2015lya}. The charges of fermions under this $U(1)$ symmetry are 
\begin{align}
X(\psi_a)=-X(\chi_a)=2~,\quad 
X(\chi_f)=X(\psi_c)=X(\psi_w)=1~,\quad 
X({\rm SM})=0~,
\end{align}
which distinguishes the exotic fermions from the SM fermions and thus can stabilize the dark matter. However, in our model, it is broken by the $\langle \Delta \rangle$, which generates the Majorana mass of $\psi_a$.}
Considering the Higgs VEV at the electroweak scale, the conserved $U(1)$ global symmetry becomes
\begin{align}
D-(B-L)=-\frac{1}{5}\left(4\,Y+\tilde{Q}\right)~.
\end{align}
The charges of all the fermions are shown in table \ref{tab:U(1)}, where we compare it with the other three global symmetries $(D,B,L)$ and identify it as $D-(B-L)$. The only unusual charge assignment is for the Dirac fermion $\psi_c$, which carries both dark baryon number and baryon number. It was assigned because it always decays to two SM quarks and one dark quark, as well be discussed later. This conserved quantum number is important in the discussion of cosmology, as it serves as the conserved $U(1)$ global symmetry that guarantees the number densities between dark baryon and baryon to be comparable.


\section{The Dark Sector}\label{sec:DS}

In this section, we focus on the dark sector and its low energy spectrum. Models with a $Sp(4)_D$ dark color gauge group have been studied in \cite{Kulkarni:2022bvh,Zierler:2022uez,Dengler:2024maq} but focus on the dark pions as dark matter candidates. In our model, instead, we have dark baryons as our dark matter, whose stability is protected by the dark baryon number $U(1)_D$. Our goal is to identify the lightest dark hadron, including dark baryons with $D=1$ and dark mesons with $D=0$.

To analyze the dark hadrons, we need to start with the dark quarks. There are two types of dark quarks in our model, $\chi_{a,i}$ and $\chi_{f,i}$, each with three generations $i=1,2,3$. They are of two different representations under the $Sp(4)_D$ dark color group, $\chi_{f,i}$ for the fundamental and $\chi_{a,i}$ for the antisymmetric. They all get current masses from the scalar VEVs as we introduced in section~\ref{sec:Model}. $\chi_{f,i}$ receive Dirac masses with a hierarchical spectrum similar to the SM quark sector. On the other hand, $\chi_{a,i}$ receive Majorana masses with a quadratically hierarchical spectrum. As already shown in section \ref{sec:Spectrum}, we expect that only the first generation of dark quarks, $\chi_{a,1}$ and $\chi_{f,1}$, are light enough with masses below the confinement scale $\Lambda_{\rm DC}$. In this section, we begin by analyzing the hadron spectrum in the massless limit, which has been extensively studied on the lattice in \cite{Bennett:2017kga,Bennett:2019jzz,Bennett:2019cxd,Bennett:2020qtj,Bennett:2022yfa,Bennett:2023mhh,Bennett:2023qwx}. Next, we discuss the realistic hadron spectrum under our benchmark scenario and highlight the relevant states in the model.

\subsection{Dark hadron spectrum in the massless limit}

\begin{figure}[tbp]
\centering
\includegraphics[width=1.0\textwidth]{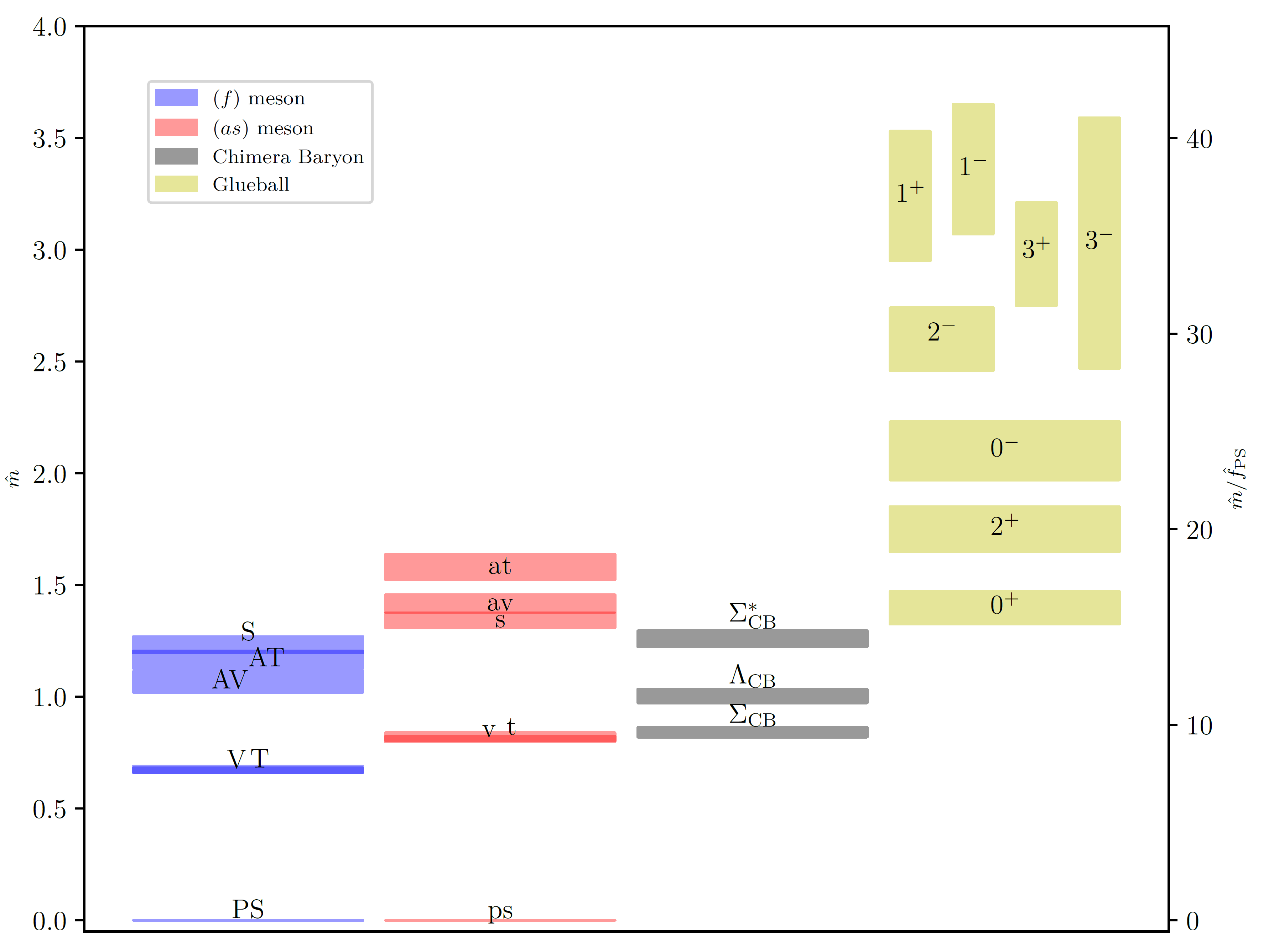}
\caption{\label{Hadron_Spectrum} The low energy spectrum of the $Sp(4)$ dark QCD secter in the massless limit. From left to right are the spectrums of mesons (fundamental/antisymmetric quarks), chimera baryons, and glueballs. The PS, V, T, AV, AT, S denote the pseudoscalar, vector, tensor, axial-vector, axial-tensor and scalar mesons of fundamental dark quarks, while letters in lowercase represents mesons of antisymmetric dark quarks. The chimera baryons are denoted following analogy with QCD baryons. The glueball states are labeled by their $J^P$ properties. The left-hand axis shows the masses in units of the gradient-flow scale and the right-hand axis shows the masses in units of decay constant of the fundamental pseudoscalar meson. This summary plot is taken from Ref.~\cite{Bennett:2023mhh}.}
\end{figure}

In the massless dark quark limit, the system features an enhanced global symmetry of $SU(6)_f \times SU(3)_a$, where the subscripts denote two different representations. Once the dark color becomes strongly coupled and the dark quark condensate forms, the global symmetry undergoes a spontaneous symmetry breaking to the subgroup $Sp(6)_f \times SO(3)a$, giving rise to fourteen (five) massless Nambu-Goldstone bosons from fundamental (antisymmetric) fermions, i.e., dark pions. On top of the pseudoscalar mesons, there are also vector, tensor, axial-vector, axial-tensor, and scalar mesons, whose masses are of the order of the dark color confinement scale $\Lambda_{\rm DC}$. Depending on the quark constituents, the spectrum can be very different, as computed using lattice simulations \cite{Bennett:2019cxd,Bennett:2023qwx} and shown in figure \ref{Hadron_Spectrum}.

One important different between a strongly coupled theory with a $Sp(4)_D$ gauge group and the SM QCD with $SU(3)_C$ gauge group is that, due to the pseudo-real nature of fundamental representation, the meson multiplets are enlarged with additional states behaving as diquarks $\chi_f \chi_f$, which carries dark baryon number $D=1$. These diquark states, according to our definition, should be considered dark baryons rather than dark mesons. Carrying a dark baryon number $D=1$, their stability is naturally protected, and they can therefore serve as dark matter candidates.

Next, we move on to the spectrum of common dark baryon states. Depending on the representations of their constituents, there are three types of baryons: fundamental-only, antisymmetric-only, and chimera baryons composed of mixed representations \cite{Ayyar:2018zuk}. Among them, the antisymmetric-only baryons are expected to be unstable due to the nature of a real representation. The fundamental-only baryons, which are common in $SU(N)$ theories, are, however, also unstable in $Sp(2N)$ theories because a baryon of $2N$ quarks with $D=N$ can still decay into $N$ diquarks with $D=1$ we just mention. That is, the baryon of this type is no longer protected by the baryon number due to the existence of $D=1$ diquarks. The last one, the chimera baryons, which are fermionic bound states composed of $\chi_a \chi_f \chi_f$ with dark baryon numbers $D=1$, is also stable dark baryon states in the theory, which is the other dark matter candidate. The chimera baryon spectrum has been derived on the lattice using quenched approximation in \cite{Bennett:2023mhh}. The results, with quarks in the massless limit, have been summarized in figure \ref{Hadron_Spectrum}, combining the spectrum of mesons \cite{Bennett:2019cxd} as well as glueballs \cite{Bennett:2020qtj}, which are bound states of pure gluons.

From the plot, it appears that the lightest baryon state is the Chimera Baryon $\Sigma_{\rm CB}$, whose notation is analogous to the $\Sigma$ baryons in QCD, by conjugating the antisymmetric dark quark with the strange quark and the fundamental dark quark with the up and down quarks. However, as mentioned earlier, there are also diquark states in the meson sector that carry a dark baryon number $D=1$. Therefore, they are stable and can also serve as dark matter. Taking these states into account, the lightest dark baryon is actually the diquark vector boson $\rho$ composed of two fundamental dark quarks, which is responsible for the observed dark matter.

\subsection{A realistic spectrum with massive dark quarks}

Next, we take into account the current masses of dark fermions, which were already discussed in section~\ref{sec:Yukawa} and shown in table \ref{tab:spectrum}, to obtain a realistic hadron spectrum. In our setup, there is one $\chi_a$ and one $\chi_f$, both with masses $\sim 0.1$ GeV right below $\Lambda_{\rm DC}\sim 1$ GeV, which together can form the various baryons discussed in the previous subsection. Most importantly, these dark baryons have their masses set by the dark color confinement scale, $\Lambda_{\rm DC}$. Note that if all dark quarks acquire masses above $\Lambda_{\rm DC}$, the dark baryon masses would instead be primarily determined by the dark quark masses originating from scalar VEVs instead of $\Lambda_{\rm DC}$, thereby spoiling the explanation for the comparable masses between dark baryons and ordinary baryons. Therefore, the presence of light dark quarks is an essential ingredient of this class of dark sector models.

Now with only one $\chi_a$ and one $\chi_f$ dark quark, the system features a decreased global symmetry of $SU(2)_f$ which is broken down to $Sp(2)_f$ after the dark quark condensate. Since there are no broken generators, there are no light dark pions as expected in one-flavor theory \cite{Creutz:2006ts,Francis:2018xjd}. Such a theory can have very different behaviors as multi-flavor theory and a dedicated lattice study is required to get a solid conclusion, which however is not available. In this study, we will try to extract the information from different study for a rough estimation.

As our goal is to identify the lightest hadron, from the spectrum in figure \ref{Hadron_Spectrum} we can conclude that the mesons of fundamental quarks are likely to be lighter than mesons of antisymmetric quarks and play rather important roles in dark sector phenomenology. Therefore, we will focus on the meson spectrum of fundamental quarks. Lattice studies based on $Sp(4)$ gauge theory with $N_f=1+1$ fundamental
quarks have been studied in \cite{Kulkarni:2022bvh,Bennett:2023rsl}. In the second quark decouple limit, the theory is similar to the $SU(2)$ gauge theory with $N_f=1$ \cite{Francis:2018xjd} which shares the same symmetry structure. From these lattice studies, a similar conclusion is stated that there are two relevant hadron states at the low energy, the pseudoscalar meson $\eta$ and the vector meson $\rho\,(\omega)$.

Start with the pseudoscalar meson $\eta$, analogous to the $\eta'$ meson in QCD, which is the Goldstone boson of anomalous $U(1)_A$ symmetry. The flavor singlet state is not shown in figure \ref{Hadron_Spectrum} but the dedicated lattice result is presented in \cite{Bennett:2023rsl,Francis:2018xjd}, showing that the $\eta$ meson is the lightest hadron in the one-flavor theory. In this work, we adapt the result from \cite{Francis:2018xjd} which gives the values of masses and decay constants in term of $\Lambda_{\rm DC}$.

The $\rho$ vector mesons, the second lightest states in the theory,\footnote{The lattice result showing that the $\rho$ vector mesons are heavier than the $\eta$ meson conflicts with the observed QCD spectrum, suggesting that the heaviness of the $\eta'$ meson in QCD might originate from the contribution of the strange quark.} form a triplet under unbroken $Sp(2)_f$. As the dark baryon number $U(1)_D$ is a subgroup of $Sp(2)_f$, decomposing the $\rho$ meson triplet under their $U(1)_D$ charge we found
\begin{align}\label{rho}
\rho=
\begin{pmatrix}
\rho^+    \\   \rho^0(\equiv\omega)    \\   \rho^- \\
\end{pmatrix}\sim
\begin{pmatrix}
~\chi_f\chi_f~\\\chi_f^c\chi_f\\ \chi_f^c\chi_f^c \\
\end{pmatrix}~\text{with}~
D=
\begin{pmatrix}
+1    \\   0    \\   -1 \\
\end{pmatrix}~.
\end{align} 
The $\rho^\pm$ with nonzero baryon number $D=\pm1$ are actually vactor diquark baryon states we have mentioned. To better distinguish these states for the following analysis, we refer $\rho^\pm$ states as $\rho$ diquark baryon but renamed the $\rho^0$ as $\omega$ meson, analogous to the iso-singlet meson in QCD. In the following discussion, unless otherwise specified, we will simply use $\rho$ to denote the $\rho^+$ dark baryon, which constitutes the dark matter.

So far, we have identified the two most relevant dark hadrons: (1) the lightest dark hadron with $D=0$, the pseudoscalar meson $\eta$, and (2) the lightest dark hadron with $D=1$, the vector baryon $\rho$, which are the same as the minimal theory of dark baryons \cite{Francis:2018xjd} based on the $SU(2)$ dark color gauge group. Although a richer dark hadron spectrum arises in the $Sp(4)$ dark color theory due to the presence of a light $\chi_a$, such as the chimera baryons, these states are irrelevant for the following discussion. This also implies that a light $\chi_a$ is not a necessary component of a successful model, although we retain the possibility that these additional states could play a role in cosmology and phenomenology.

To explain the dark matter-baryon coincidence, with the number densities satisfying $n_D = c\,n_B$, we require the mass of $\rho$ to be $m_\rho = (5/c) \times m_B$. In this study, we choose $c = 3$ as our benchmark, which will be justified in section \ref{sec:Asy}. The desired mass then becomes $m_\rho = (5/3) \times m_B \sim 1.6$ GeV. From a lattice study \cite{Francis:2018xjd}, we find that it can be satisfied with $m_{f,1}=0.1$ GeV and $\Lambda_{\rm DC}=0.75$ GeV, giving $m_\rho \sim 2.2 \,\Lambda_{\rm DC} \sim 1.6$ GeV and $m_\eta \sim 1.4 \,\Lambda_{\rm DC} \sim 1.0$ GeV. The decay constant are also derived as $f_\rho \sim 550$ MeV and $f_\eta \sim 150$ MeV. The desire dark color confinement scale $\Lambda_{\rm DC}=0.75$ GeV can be achieved with certain choice of $\alpha_4'$ as discussed in section~\ref{sec:Spectrum}. We will use these benchmark values for the following study of dark sector cosmology and phenomenology.


\section{Cosmology}\label{sec:COS}

In this section, we describe a successful cosmological evolution. The idea follows the standard framework of Asymmetric Dark Matter (ADM) models \cite{Spergel:1984re,Nussinov:1985xr,Gelmini:1986zz,Barr:1990ca,Barr:1991qn,Kaplan:1991ah,Gudnason:2006ug,Gudnason:2006yj,Kaplan:2009ag,Davoudiasl:2012uw,Petraki:2013wwa,Zurek:2013wia}, involving both the generation of asymmetries and the annihilation of the symmetric component. Within the framework of dark grand unified theories, many of the required ingredients are naturally provided. We also discuss the stability of the dark baryon in this section.

\subsection{Generation of the dark matter and baryon asymmetries}\label{sec:Asy}

The generation of the asymmetries can follow two different paths. One can first have an asymmetry created in one sector and then communicate it to the other through some transfer mechanisms. Alternatively, the dark baryon and baryon asymmetry can be generated simultaneously. Both cases can fit well within our model.

For the first case, known as "Sharing," the new gauge interactions at the GUT/DU scale can play a role in transferring the primordial asymmetry. Besides, one can easily embed an additional gauge group, such as a horizontal gauge group, and use the new sphaleron processes to realize the transfer.

In this study, we focus on the second realization, known as "Cogenesis," where the asymmetry of the visible sector and the dark sector is generated through the same process. To realize this scenario, the three Sakharov conditions \cite{Sakharov:1967dj} must be satisfied, including
\begin{enumerate}
  \item Dark baryon and baryon/lepton number violation
  \item Charge conjugation (C) and charge conjugation parity (CP) violation
  \item Out of thermal equilibrium process
\end{enumerate}
In the dark grand unified theory, there are many processes that violate both the dark baryon and baryon/lepton numbers, satisfying the first condition. With new couplings among the new states, especially those involving three generations of new fermions, additional C/CP violation sources are also expected. If the process occurs slowly and remains out of thermal equilibrium, we can then achieve all the required conditions.

\begin{figure}[tbp]
\centering
\includegraphics[width=0.9\textwidth]{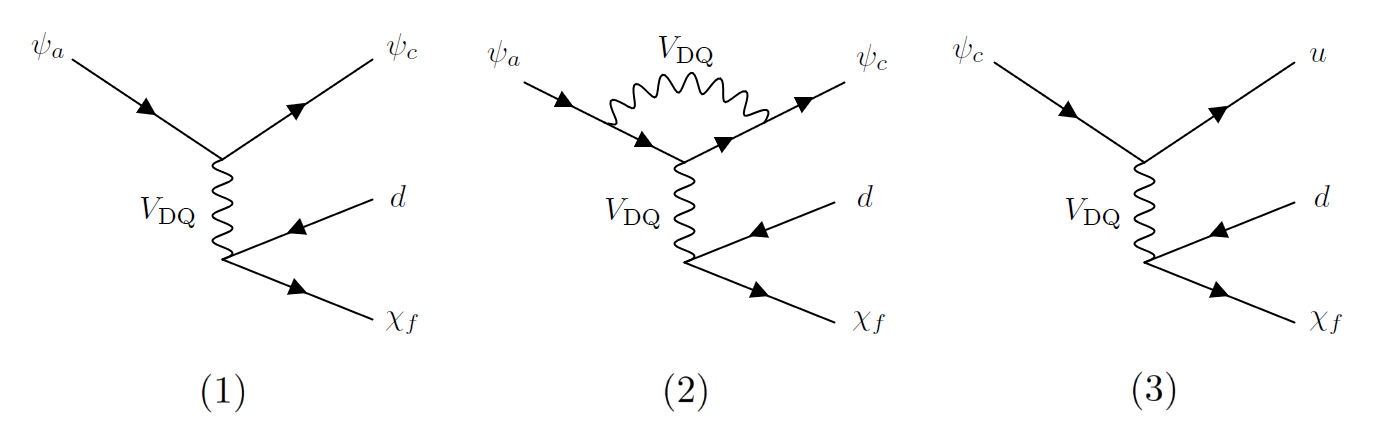}
\caption{Feynman diagrams of the relevant decay processes for the generation of the dark matter and baryon asymmetries. From left to right: (1) the tree-level diagram of $\psi_a$ decay, (2) the loop-level diagram of $\psi_a$ decay, and (3) $\psi_c$ decay into SM fermions and dark quarks.\label{fig:psidecay}}
\end{figure}

In this work, we use the out-of-equilibrium decay of $\psi_{a,i}$ to source the asymmetries, analogous to the dark-baryo-genesis process in \cite{Murgui:2021eqf}. The $\psi_{a,i}$ particles must decay through one of the bosons that introduce species-changing interactions, and the lightest one is likely to dominate the decay channels. Following the spectrum discussed in section \ref{sec:Spectrum}, the scalar bosons are heavier than the vector bosons. Therefore, the lightest boson capable of mediating $\psi_{a,i}$ decay is $V_{\rm DQ}$, which couples to the dark-colored particles and colored particles. The relevant interaction terms are given by 
\begin{align}
{\cal L}_{\text{DQ}}= 
\frac{g_{7}}{\sqrt{2}}\,V_{\rm DQ}^{\mu} (U_{ca}\,\bar{\psi}_{c,i}\gamma_\mu\psi_{a,i}
+U_{df}\,\bar{d}_i^c\gamma_\mu \chi_{f,i}
+U_{uc}\,\bar{u}_i^c\gamma_\mu \psi_{c,i}^c)\,,
\end{align}
which allows the decay $\psi_a \to \psi_c\, V_{\rm DQ}$. The unitary matrices $U_{ab}$, where $a$ and $b$ denote different species, are the product of unitary matrices shown in section \ref{sec:Yukawa}, analogous to the CKM matrix in the SM weak interactions. As discussed in section \ref{sec:Spectrum}, the vector bosons are heavier than the fermions. Therefore, the $V_{\rm DQ}$ in the final state must be off-shell, which can decay into light fermions, such as $V_{\rm DQ} \to d\,\chi_f$, resulting in an overall three-body decay $\psi_a \to \psi_c\, d\,\chi_f$, as shown in (1) of figure \ref{fig:psidecay}.

The $\psi_c$ field in the final state is still unstable and can subsequently decay into SM fermions and light dark quarks, also through $V_{\rm DQ}$, as $\psi_c \to u\,d\,\chi_f$ shown in (3) of figure \ref{fig:psidecay}. The complete decay chain is then given by
\begin{align}\label{decayprocess}
\psi_a\to \psi_c \,d\,\chi_f \to u\,d\,d\,\chi_f\,\chi_f~.
\end{align}
The decay chain features dark baryon number violation and baryon number violation, satisfying the first Sakharov condition. Moreover, it satisfies $\Delta(D-B) = 0$, which ensures that the number densities generated in the two sectors are the same.

The second condition requires the CKM-like matrices $U_{ab}$ with non-vanishing CP phases in the relevant vertices, such as $U_{ca}$ in the decay process we use. The phases originate from the complex Yukawa coupling matrices in eq. \eqref{Yukawa}. CP-violating decays can then arise from the interference of multiple $1 \to 3$ diagrams, such as (1) and (2) in figure \ref{fig:psidecay}, with the one-loop penguin diagram (2) used as an example. Through processes that involve fermions of all three generations, this can result in a nonzero $\epsilon_{\rm CP}$.\footnote{A similar leptogenesis process through three-body decay, with detailed calculations of $\epsilon_{\rm CP}$, has been studied in \cite{Hambye:2001eu, Borah:2020ivi, Dominguez:2023rnx}. In this study, we will leave $\epsilon_{\rm CP}$ as a free parameter.}

Last, we turn to the third requirement for successful asymmetry generation. We focus on the first step of the decay $\psi_a \to \psi_c \,d\,\chi_f$. Assuming this to be the leading decay channel, the decay rate of $\psi_a$ is then given by
\begin{align}
\Gamma_{a}\sim \frac{1}{384\pi^3}\frac{M_a^5}{f^4}\sim  7\times 10^{32} ~s^{-1}~.
\end{align}
For the decay to be out of thermal equilibrium, the lifetime needs to be greater than the age of the universe at the temperature right below the mass of the decaying particle, which can be expressed in terms of  the Hubble parameter
\begin{align}
H|_{T\,\lesssim\,M_a}\sim 1.7\sqrt{g_*}~\frac{M_a^2}{M_{\rm Planck}}\sim 3.5\times 10^{35} ~s^{-1}~,
\end{align}
where $g_*=334.25$ is the the effective degrees of freedom at $T\,\lesssim\,M_a$ following the spectrum in table \ref{tab:spectrum}. We can see that the lifetime of $\psi_a$, $\tau_{a} = \Gamma_{a}^{-1}$, although very short, is still longer than the age of the universe, $t \sim H^{-1}$, allowing the decay to be out of thermal equilibrium.\footnote{In this work, we focus on cogenesis triggered by three body decay due to the spectrum with $M_a<M_{DQ}$. One can also consider the scenario with $M_a>M_{DQ}$ and cogenesis through two body decay, which would require $M_a>10^{16}$ GeV to be out of thermal equilibrium.}

With all the quantities, we can estimate the amount of asymmetry generated through this process, which is given by
\begin{align}
Y_{D}=Y_{B} \sim \epsilon_{\rm CP}\times \frac{T_{a,\rm reheat}}{M_a}\sim 0.027~\epsilon_{\rm CP}~,
\end{align}
where $T_{a,\rm reheat}$ is the reheat temperature of the $\psi_a$ decay given by
\begin{align}
T_{a,\rm reheat}\sim \left(\frac{45}{16\pi^3g_*}\right)^{1/4}\sqrt{\Gamma_{a}M_{\rm Planck}}\sim 9.6\times 10^{12} ~{\rm GeV}~.
\end{align}
The potential dilution from early matter domination and the washout effects from scattering processes are not taken into account. We find that the asymmetries generated in the two sectors are identical due to the conservation of $D - B$ encoded in the processes, and only a small $\epsilon_{\rm CP} \sim \mathcal{O}(10^{-8})$ is required to reproduce the observed abundances. However, the relation $Y_{D}=Y_{B}$ is still violated by the effect of the electroweak sphaleron, which reduces $Y_{B}$ by $\sim 2/3$ of the net $Y_{B-L}$ asymmetry generated at the DU scale. This results in the final $Y_{D}\sim 3\times Y_{B}$ \cite{Harvey:1990qw}, which explains the use of $n_D=3\,n_B$ in our benchmark.

\subsection{Stability of dark baryon dark matter}

Since the dark baryon number can be violated, the lightest dark baryon, in principle, can decay through a heavy mediator at the DU scale or the GUT scale. If this is the case, we should check whether its lifetime satisfies the current constraints, which are given by indirect detection \cite{Papucci:2009gd,Cirelli:2009dv,Essig:2013goa}, with the bound $\tau_D > 10^{28}~\text{s}$. Fortunately, in our model, the dark vector baryon decay is kinematically forbidden due to its special quantum numbers, which can be verified through the corresponding conservation laws.

To examine this, first note that $\Delta(D - B + L) = 0$ are satisfied in all the interactions we introduce, implying that, for the dark baryon decays, there must be at least one lepton or one anti-baryon in the final state. Both a lepton and an anti-baryon are fermions, while the dark baryon in our model is a vector boson. Therefore, another fermion with $D - B + L = 0$ needs to be present in the final state. However, the only light fermion that carries a vanishing $D - B + L$ charge is the dark quark $\chi_a$. Therefore, unless there exists a light fermionic dark hadron composed of $\chi_a$, which is not the case in our setup, the decay of the dark $\rho$ baryon is kinematically forbidden. In other words, the absence of a light $D - B + L = 0$ fermion guarantees the stability of the lightest $D - B + L = 1$ boson - the dark $\rho$ baryon in our model.\footnote{We are aware that lepton number can be violated in the SM due to the presence of neutrino masses. However, the problem persists if lepton number violation occurs only by two units, i.e., $|\Delta L| = 2$, as in typical Majorana neutrino scenarios.}


\subsection{Annihilation of the symmetric abundances}

Following the discussion in subsection \ref{sec:Asy}, after the small amount of asymmetry between dark baryons and anti-dark baryons is generated, we also need the remaining symmetric abundance to annihilate and transfer the extra entropy to the SM sector to satisfy the current measurements. This process involves two essential steps.

\begin{figure}[tbp]
\centering
\includegraphics[width=1.0\textwidth]{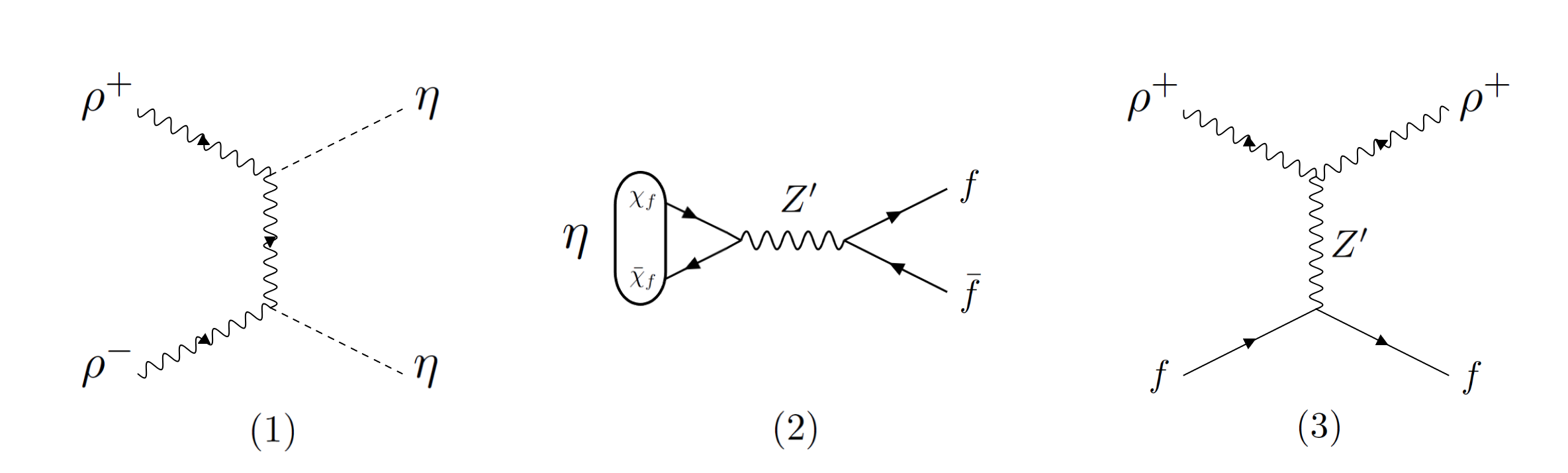}
\caption{Relevant processes among dark hadrons and SM fermions, including (1) dark $\rho$ baryons annihilation (2) dark $\eta$ meson decays (3) interactions between dark $\rho$ baryons and SM fermions.\label{fig:decay}}
\end{figure}

First, the dark baryons and anti-dark baryons annihilate into the lightest hadrons in the dark QCD sector. In our case, this process is $\rho^+ \rho^- \to \eta \eta$, as shown in (1) of figure~\ref{fig:decay}, which is kinematically allowed according to the lattice results \cite{Francis:2018xjd}. The annihilation cross section is determined by the dark color dynamics. Although the interactions among hadrons are non-perturbative and therefore can only be obtained through first-principles calculations. A generic estimate of the annihilation cross section, $\langle \sigma v \rangle \sim 1/\Lambda_{\rm DC}^2 \sim 10^{-17}~\text{cm}^3\,\text{s}^{-1}$ with $\Lambda_{\rm DC} \sim 1$ GeV, indicates that the annihilation process is certainly fast enough to deplete the symmetric abundance into dark $\eta$ mesons~\cite{Lin:2011gj}.


Next, if these dark $\eta$ mesons, which carry the entropy of the symmetric abundance, are stable, they could overclose the universe. To satisfy current observations, they must decay into the SM sector before the Big Bang Nucleosynthesis (BBN) epoch \cite{Jedamzik:2006xz}, which requires a lifetime $\tau_{\eta} < 1~\text{s}$. In our model, the decay is mediated by the $Z'$ boson associated with the broken $U(1)'_5$ symmetry, leading to a pair of SM fermions, as shown in (2) of figure~\ref{fig:decay}. After integrating out the $Z'$ boson, the effective interaction terms are given by
\begin{align}
{\cal L}_{\eta} = c_\chi\left(f_\eta\partial^\mu \eta\right)\frac{g_{Z'}^2}{M_{Z'}^2}c_f\left(\bar{f}\gamma_\mu\gamma_5 f\right)
=c_f\frac{\partial^\mu\eta}{f_a} \left(\bar{f}\gamma_\mu\gamma_5 f\right)~\text{ with }~ f_a=\frac{c_{f'}^2f'^2}{c_\chi f_\eta}~,
\end{align}
which behaves like a typical axion-like particle (ALP) \cite{Dolan:2014ska,Dobrich:2018jyi,Bauer:2020jbp,Bauer:2021mvw,DallaValleGarcia:2023xhh,Ovchynnikov:2025gpx} that couples dominantly to fermions with flavor-diagonal couplings due to the properties of the $Z'$ boson. The coefficients $c_\chi$ and $c_f$ are determined by the charges of $\chi_f^{(c)}$ and the SM fermions, as already listed in Eq.~\eqref{Q'5}, following the relation $c_f = Q'_5(f) + Q'_5(f^c)$. We find that $c_\chi = (3 + c)/4$, which has the same magnitude as $c_{f'}$. Furthermore, the coefficients for SM fermions satisfy $-c_u = c_d = c_\ell = c_\nu = 1$, implying a universal coupling structure.

Considering the benchmark point with a confinement scale $\Lambda_{\rm DC} = 0.75$ GeV, which gives $m_\eta = 1.4\,\Lambda_{\rm DC} = 1.0$ GeV and $f_\eta = 0.2\,\Lambda_{\rm DC} = 150$ MeV, we can estimate the lifetime of the dark $\eta$ meson. We begin with the leptonic decay, where $\eta \to \mu^+ \mu^-$ is the dominant decay channel. The corresponding decay width is given by
\begin{align}\label{rho_decaywidth}
\Gamma_{\eta\to\mu^+\mu^-}
\sim\frac{1}{2\pi}\frac{m_{\eta}m_\mu^2}{f_a^2}
\sim 0.4 ~s^{-1}\left(\frac{10^5 \text{ GeV}}{f'}\right)^4\left(\frac{\Lambda_{\rm DC}}{0.75 \text{ GeV}}\right)^3
~,
\end{align}
where the dependence on the two relevant scale parameters is shown. The hadronic decay channels are more complex. Following the analysis in \cite{DallaValleGarcia:2023xhh}, we expect that for $m_\eta = 1.0$ GeV, the dominant hadronic decay channel is $\eta \to \eta_{\rm SM}\pi\pi$, whose decay width is a few times larger than $\Gamma_{\eta \to \mu^+ \mu^-}$. Taking both contributions into account, we find that the lifetime of the dark $\eta$ meson is less than one second, thus satisfying the constraint from BBN.

From the analysis above, we can also see that the scale $f'$ cannot be arbitrarily high. Otherwise, the dark $\eta$ meson would be too long-lived and would violate the BBN constraint. Therefore, a rough upper limit of $f' \lesssim 10^5~\text{GeV}$ is required for successful cosmology, leaving a finite parameter space that may lead to testable predictions.


\section{Phenomenology}\label{sec:Pheno}

The model features a GeV-scale dark QCD-like sector with dark baryons and dark mesons that can be probed experimentally, similar to \cite{Murgui:2021eqf,Francis:2018xjd}. Besides, the presence of the $Z'$ boson and its crucial role in the framework distinguish our phenomenology from others. Moreover, the $Z'$ boson itself could potentially be detected at colliders, along with other exotic states predicted by the model. Finally, we highlight the potential importance of dark matter self-interactions, which could have a significant impact on astronomy.

\subsection{Searches of dark mesons}

We begin with the lightest particle in the dark sector, the pseudoscalar meson $\eta$, whose properties were also discussed in the previous section. Since it couples to SM fermions through the $Z'$ boson, it behaves like an axion-like particle, coupling predominantly to fermions with flavor-diagonal and universal interactions \cite{Dolan:2014ska,Dobrich:2018jyi,Bauer:2020jbp,Bauer:2021mvw,DallaValleGarcia:2023xhh,Ovchynnikov:2025gpx}. In this subsection, we follow the results presented in \cite{Ovchynnikov:2025gpx}.

Considering the dark $\eta$ meson with $m_\eta = 1.0$ GeV and $f_\eta = 150$ MeV, one strong constraint come from the B-meson decay measurement by LHCb \cite{LHCb:2015nkv,LHCb:2016awg}, requiring the scale  $f' \gtrsim 600$ GeV. The strongest existing constraint, however, comes from an old beam-dump experiments. The lower bound $f' \gtrsim 2000$ GeV is set by CHARM collaboration \cite{CHARM:1985anb}. The bound can be pushed forward for an order of magnitude in the future experiments, such as SHiP \cite{SHiP:2015vad} and MATHUSLA \cite{Curtin:2018mvb}, which will probe $f'$ values as high as $\sim 10^4$ GeV, close to the upper bound set by the BBN constraint.

\subsection{Direct detection of dark baryons}

Also through the $Z'$ boson, the dark baryon $\rho$ can interact with the SM quarks and thus nucleons as shown in (3) of figure \ref{fig:decay} with the cross section given by \cite{Lin:2019uvt}
\begin{align}
\sigma_n \sim \frac{m_{\rho n}^2}{\pi}\left(\frac{g_{Z'}^2}{M_{Z'}^2}\right)^2
\sim 4.5\times 10^{-47}~{\rm cm^2}\left(\frac{m_{\rho n}}{0.6 \text{ GeV}}\right)^2\left(\frac{10^5 \text{ GeV}}{f'}\right)^4~,
\end{align}
where $m_{\rho n}$ is the reduced mass of the dark baryon and SM nucleon, $m_{\rho n}={m_\rho m_n}/({m_\rho+m_n})$, which is $0.6$ GeV under our benchmark.

The current constraint on the direct detection, considering a $1.6$ GeV dark matter, comes from the argon-based detector, DarkSide-50 \cite{DarkSide:2018bpj,DarkSide-50:2022qzh}, putting a 90\% C.L. exclusion limit for the spin-independent cross section at $\lesssim 10^{-41}~{\rm cm^2}$, corresponding to the bound $f' \gtrsim 5000$ GeV, which is one order of magnitude weaker than the bound derived from dark meson searches. The successor, DarkSide-20k \cite{DarkSide-20k:2024yfq} and DarkSide-LowMass \cite{GlobalArgonDarkMatter:2022ppc}, can put the exclusion limit for the cross section further down for another two orders of magnitude. The future detector could ultimately reach the neutrino-fog at $\sim 10^{-45}~{\rm cm^2}$ \cite{OHare:2021utq}, putting the constraint on the scale $f'$ up to $5\times 10^4$ GeV.


\subsection{Collider signatures and flavor observables}

Looking back at our spectrum in table \ref{tab:spectrum}, there are several new states that directly couple to the SM fermions and are light enough to have impacts on LHC searches (below $10$ TeV) or flavor observables (below $10^{3}$ TeV).

Start with the $Z'$ boson, which mediates important interactions between dark sector and the SM sector. Besides indirect bound on the scale $f'$, we can also look for the $Z'$ boson directly. The $Z'$ boson couples to both quarks and leptons, allowing the searches in the simplest dilepton channel, $pp \to Z' \to \ell^+\ell^-$. The current constraint from LHC searches already places a bound of $M_{Z'} \gtrsim 5$ TeV \cite{ATLAS:2019erb,CMS:2021ctt}. Taking a typical gauge coupling strength $g_{Z'} \sim 0.3$, the collider searches bound would have given the strongest constraint on the scale as $f' > 1.3 \times 10^4$ GeV. With a 100 TeV collider in the future, we can look for ${Z'}$ bosons as heavy as $40$ TeV, which can examine the scale $f'$ up to $10^5$ GeV, the entire parameter space of the theory. If the $Z'$ boson is detected, we can then check if it actually couples to the dark sector as proposed by measuring its decay width.

In principle, flavor observables can probe even higher scales compared to collider searches. Unfortunately, since the $Z'$ boson in our setup couples to the SM fermions in a flavor-universal and flavor-diagonal manner, there are no flavor-changing neutral currents, and thus the strong constraints from flavor physics do not apply here. Nevertheless, since the $U(1)_5$ charges are assigned by hand, one could instead consider flavor non-universal charge assignments \cite{Davighi:2023iks,Chung:2021ekz,Chung:2021xhd}, which might provide hints about the origin of flavor and lead to interesting flavor observables that could be tested.

In addition to the $Z'$ boson, another state potentially detectable at colliders is the second Higgs doublet, as in the standard Type-II 2HDM \cite{Branco:2011iw}. The additional heavy Higgs bosons receive strong constraints from both direct searches and flavor observables \cite{Kling:2020hmi,Li:2024kpd}. In particular, the charged Higgs boson mass below $800$~GeV is strongly constrained by the precision measurements of flavor-changing process $b\to s\gamma$ \cite{Misiak:2017bgg,HFLAV:2019otj}, requiring a nontrivial scalar potential to separate the SM Higgs boson and the heavy Higgs bosons.

The other fields that might be around an accessible scale are the $\Phi'$ scalar field, whose VEV is directly responsible for the scale $\Lambda_5$. After decomposition, the scalar field is separated into $\Phi'_d$, $\Phi'_c$, and $\Phi'_w$. The $\Phi'_d$, especially the singlet component, must be around the scale $\Lambda_5$. However, as it only couples to the dark sector, the effect has less phenomenological relevance. On the other hand, if $\Phi'_c$ or $\Phi'_w$ fields, which couple to dark and visible sector, are light enough, they could introduce interesting signatures in flavor observables, which have been studies under the framework of a flavored dark sector \cite{Kile:2011mn,Batell:2011tc,Kamenik:2011nb,Agrawal:2011ze,Lopez-Honorez:2013wla,Kile:2013ola,Batell:2013zwa,Kile:2014jea,Renner:2018fhh}.

\subsection{Dark Matter self-interactions}

One important feature of the dark baryon model is a strong dark matter self-interaction, which is non-perturbative with the calculations relying on first-principles methods. In this study, we estimate the scattering length to be determined by the confinement scale $\Lambda_{\rm DC}$, $a\sim \Lambda_{\rm DC}^{-1}$. With $m_{\rho}=2.2\,\Lambda_{\rm DC}$, we get the cross section per unit mass for dark baryon self-interactions given by
\begin{align}
\frac{\sigma_D}{m_D} \,\sim\, \frac{\pi\,\Lambda_{\rm DC}^{-2}}{m_\rho} \, = 7\times 10^{-4} \left(\frac{0.75 \text{ GeV}}{\Lambda_{\rm DC}}\right)^3 ~{\rm cm^2/g}~.
\end{align}
The value is much smaller than the common constraint extracted from the observation of Bullet Cluster \cite{Markevitch:2003at, Randall:2008ppe, Robertson:2016xjh}, which requires ${\sigma_D}/{m_D} \lesssim 1~{\rm cm^2/g}$,  and even a much stronger constraint from cluster strong lensing \cite{Andrade:2020lqq}, which requires ${\sigma_D}/{m_D} \lesssim 0.1~{\rm cm^2/g}$.

The scattering length can be much greater if there is resonant enhancement due to the presence of bound states. It is the case in the SM QCD, where the scattering length of neutron-proton scattering is enhanced dramatically to $a\sim 24\,{\rm fm} = (8 \,{\rm MeV})^{-1}$, much greater than the generic estimation by the pion mass $m_\pi^{-1}\sim (140 \,{\rm MeV})^{-1}$, due to the deuteron bound state. Whether the same effect occurs in the dark QCD sector can only be checked by first-principles calculations, which is beyond the scope of this study.

If there is no resonant enhancement, the constraint on the dark matter scattering cross section is easily satisfied due to lack of long range forces. However, this only applies to our specific dark quark spectrum, which is pionless due to lack of approximate global symmetry. If one or more dark quarks, either $\chi_a$ or $\chi_f$, has a mass below $\Lambda_{\rm DC}$, the cross section might be enhanced due to the existence of light dark pions $\pi'$. A sizable dark matter self-interactions with scattering length ($a\sim \mathcal{O}(10)\,{\rm fm}$) can leave an imprint on the galaxy-scale, potentially relieving the long standing core-cusp problem \cite{Spergel:1999mh,Dave:2000ar,Kaplinghat:2015aga,Tulin:2017ara,Chu:2019awd,Cline:2013zca,Cline:2022leq}. As the generic scale of the required cross section matches the scale of the dark sector needed to address the dark matter-baryon coincidence problem, these astronomical anomalies might already provide the first hint of a GeV-scale strongly coupled dark sector. We leave a detailed study of this possibility for future work.


\section{Conclusions}\label{sec:Conclusion}


We have constructed a Dark Grand Unification model to explain the dark matter–baryon coincidence problem. With the dark color $Sp(4)_D$ and dark quarks, which automatically arise from the extension of the $SU(5)$ GUT fermion content, the dark color confinement scale can be comparable to the QCD confinement scale under some conditions, explaining the comparable masses between dark baryons and baryons. Furthermore, the model preserves a $U(1)$ global symmetry, $D - (B-L)$, which guarantees that the number densities of the two sectors should also be similar. The required cosmological process can be easily realized through the ingredients presented in the Dark GUT. With these two key features, the observed $\rho_D / \rho_B \approx 5$ can then be explained.


The dark QCD sector exhibits some interesting properties. The $Sp(4)_D$ gauge group features diquark dark baryons which are not presented in a common $SU(N)$ theory. In addition, there are three generations of dark quarks, the same as in the SM, which introduce a flavor pattern to the dark QCD sector and provide new CP violation sources relevant for the generation of asymmetry. In this work, we focus only on the specific spectrum consistent with our idea, but there could be other possibilities leading to novel and flavorful phenomenology that is worth exploring.


With a GeV-scale strongly coupled dark sector, we expect that both dark baryons and dark mesons lie within the mass range accessible to future beam-dump experiments and dark matter direct detection searches. Moreover, the $Z'$ boson, which mediates the two sectors, could also be probed in current and future collider experiments. Since the parameter space is finite, the model is fully testable, and any of these signatures could provide valuable insight into the nature of dark matter.


\acknowledgments

I thank Avik Banerjee, Florian Goertz, and \'Alvaro Pastor-Guti\'errez for many useful discussions and Shihwen Hor for the setup of the Higgs sector. I also thank C.-J. David Lin, Ed Bennett, Deog Ki Hong, Ho Hsiao, Jong-Wan Lee, Biagio Lucini, Maurizio Piai, and Davide Vadacchino for letting me use the figure in Ref.~\cite{Bennett:2023mhh}. I am also very grateful to the Mainz Institute for Theoretical Physics (MITP) of the Cluster of Excellence PRISMA+ (Project ID 390831469), for its hospitality and its partial support during the completion of this work.

\appendix

\section{The $SU(8)$ dark grand unification with $SU(3)_D$ dark QCD sector}\label{App:SU(3)}

The $SU(3)_D$ dark color gauge group is another viable and most straightforward choice, sharing the same $C_2(G) = 3$ as QCD. Following the same construction as in section~\ref{sec:Model}, we can extend the $SU(5)$ GUT gauge group to an $SU(8)$ dark GUT gauge group to incorporate the $SU(3)_D$ dark color group. For each generation, the fermion content is extended to
\begin{align}
A=\mathbf{28}=
\begin{pmatrix}
\chi_a      &   \chi_c   &  \chi_w \\
-\chi_c^T   &   u^c      &  q      \\
-\chi_w^T   &   -q^T     &  e^c    \\
\end{pmatrix}
~\text{ and }~
\bar{F}=\overline{\mathbf{8}}=
\begin{pmatrix}
\chi_f \\ d^c \\ \ell 
\end{pmatrix}~
\label{SU(8)F}
\end{align}
with four new types of fermions labeled by $\chi$ of different subscripts reflecting their quantum number. They transform under the gauge group $SU(3) \times SU(3)_C \times SU(2)_W \times U(1)_Y$ as 
\begin{align}\label{chiA}
\chi_a= (\bar{3},1,1)_0~,~\chi_c=(3,3,1)_{-\frac{1}{3}}~,~\chi_w =(3,1,2)_{\frac{1}{2}}~,~ \chi_f= (\bar{3},1,1)_0~.
\end{align}
We again find two exotic fermions, $\chi_c$ and $\chi_w$, that are charged under both the dark color and the SM gauge group, and two SM-singlet fermions, $\chi_a$ and $\chi_f$, that again can serve as dark quark candidates.

Similar to the $SU(9)$ case discussed in the main text, we introduce three additional $\overline{\mathbf{8}}$ anti-fundamental representations for each generation to cancel the quantum anomalies. Moreover, we gauge the $SU(3)'$ global symmetry among these additional $\overline{\mathbf{8}}$ fields and identify the diagonal subgroup of $SU(3) \times SU(3)'$ as the dark color gauge group $SU(3)_D$. The new fermions transform under the $SU(8) \times SU(3)'$ gauge group as
\begin{align}
\Psi = (\overline{\mathbf{8}},\overline{\mathbf{3}})=
\begin{pmatrix}
\psi_f \\ \psi_c \\ \psi_w 
\end{pmatrix}=
\begin{pmatrix}
\psi_s\oplus\psi_a \\ \psi_c \\ \psi_w 
\end{pmatrix}
\label{SU(9)F}~.
\end{align}
Together with $\chi$ in eq. \eqref{chiA}, we get new dark-colored fermions with quantum numbers
\begin{align}\label{newfermionA}
&\chi_a= (\bar{3},1,1)_0~,~ \chi_f= (\bar{3},1,1)_0~,~\chi_c=(3,3,1)_{-\frac{1}{3}}~,~\chi_w =(3,1,2)_{\frac{1}{2}}~,\nonumber\\
&\psi_s= (\bar{6},1,1)_0~,~ \psi_a= (3,1,1)_0~,~ \psi_c=(\bar{3},\bar{3},1)_{\frac{1}{3}}~,~ \psi_w =(\bar{3},1,\bar{2})_{-\frac{1}{2}}
\end{align}
under the gauge group $SU(3)_D \times SU(3)_C \times SU(2)_W \times U(1)_Y$, which is similar to the fermion content of the $SU(9)$ model in eq. \eqref{newfermion} up to some singlet fermions. We can identify pairs of Weyl fermions that can form Dirac fermions, such as $(\chi_c,\psi_c)$ and $(\chi_w,\psi_w)$. So far, the construction is largely analogous to the $SU(9)\times Sp(4)'$ model.

\subsection{Drawback 1: $SU(3)'$ Anomaly Cancellation}
Then comes the first difference and also a drawback, regarding the anomaly cancellation of the $SU(3)'$ gauge group. Since there are already eight $\overline{\mathbf{3}}$ representations under $SU(3)'$, additional fermions charged under $SU(3)'$ are required to achieve anomaly cancellation.

One possible solution is to introduce eight additional fermions, ${\chi'_f}_i$, transforming as fundamentals under the $SU(3)'$ group, thereby rendering the $SU(3)'$ sector vector-like. However, after the symmetry breaking $SU(3)\times SU(3)'\to SU(3)_D$, the resulting dark QCD sector (assuming that the Weyl fermion pairs $(\chi_c,\psi_c)$ and $(\chi_w,\psi_w)$ acquire Dirac masses and thus decouple at low energies) features a chiral fermion content with
\begin{align}\label{chiralfermion1}
\chi_a= (\bar{3},1,1)_0,~ \chi_f= (\bar{3},1,1)_0,~\psi_s= (6,1,1)_0,~ \psi_a= (3,1,1)_0,~{\chi'_f}_i=8\times ({3},1,1)_0.
\end{align}
Such a chiral dark sector may give rise to massless composite fermions after confinement \cite{Raby:1979my,Bars:1981se,Appelquist:2000qg,Csaki:2021xhi,Bolognesi:2021jzs,Smith:2021vbf,Karasik:2022gve}, in contrast to QCD-like confining theories. Therefore, it may fail to produce the desired dark baryon whose mass is set by the confinement scale.

The other choice is to add a symmetric representation ${\chi'_s}$ together with a fundamental representation ${\chi'_f}$. The dark QCD sector then has a fermion content with 
\begin{align}\label{chiralfermion2}
&\chi_a= (\bar{3},1,1)_0,~ \chi_f= (\bar{3},1,1)_0,~\psi_s= (\bar{6},1,1)_0,~ \nonumber\\
&\psi_a= (3,1,1)_0,~{\chi'_f}=({3},1,1)_0,~{\chi'_s}=(6,1,1)_0,
\end{align}
which can result in a vector-like dark QCD sector. However, the pairs of Weyl fermions that form singlets may acquire Dirac masses and be absent from the low-energy dark sector.

\subsection{Drawback 2: Fermion spectrum and $SU(3)_D$ running}

From the fermion content in eq.~\eqref{chiralfermion2}, which includes two Dirac fermions in the fundamental representation and one Dirac fermion in the two-index symmetric representation, we find that to achieve a running comparable to QCD, the desired fermion spectrum should retain the two triplets at low energies while placing the sextet at a higher scale. This setup leads to a quark spectrum analogous to that of the SM and thus a comparable beta function. However, the scalar field $\Phi=(\mathbf{8},{\mathbf{3}})$, analogous to eq.~\eqref{Phi}, which breaks the $SU(3)\subset SU(8)$ and $SU(3)'$ into diagonal subgroup $SU(3)_D$ and generates Dirac masses for $(\chi_c,\psi_c)$ and $(\chi_w,\psi_w)$, will also introduce $(\chi_a,\psi_a)$ masses through
\begin{align}\label{YukawaA}
-{\cal L}_{\rm Yukawa} = Y A \Psi \Phi^* = 
Y_{i,j} \Phi^* \left(\chi_{c,i}\psi_{c,j}+\chi_{w,i}\psi_{w,j}+\chi_{a,i}\psi_{a,j}\right)+\text{h.c.}~
\end{align}
once the $\Phi$ field acquires a non-zero VEV at the DU scale. This situation also occurs in the $Sp(4)_D$ model. However, since the two-index antisymmetric representation under $Sp(4)_D$ is real, a Majorana mass for $\psi_a$ can be written. Consequently, the dark quark $\chi_a$ can remain in the low-energy spectrum via a seesaw mechanism. In contrast, this is not possible in the $SU(3)_D$ model, where the two-index antisymmetric representation $\bar{\mathbf{3}}$ is complex. Therefore, a more elaborate construction is required, either by modifying the Yukawa couplings in eq.~\eqref{YukawaA} to make $(\chi_a, \psi_a)$ light or by adjusting the running through the remaining charged particle masses.


Although it presents drawbacks that make realizing dark baryon masses at the GeV scale challenging for our purposes, this extension still provides an elegant framework for constructing an $SU(3)_D$ dark sector, or more generally an $SU(N)_D$ dark sector, with several interesting and novel features.

First, it naturally introduces SM-singlet fermions that can serve as dark quark candidates, which can be stabilized by accidental global symmetries such as dark baryon number. The required $U(1)_{D-(B-L)}$ symmetry for asymmetric dark matter models can also arise naturally within this framework. Moreover, it introduces exotic fermions charged under both the dark color group $SU(N)_D$ and the SM color group $SU(3)_C$, which are commonly present in other dark sector models addressing the coincidence problem \cite{Bai:2013xga,Newstead:2014jva,Ritter:2022opo,Ritter:2024sqv}.

Moreover, it not only introduces a QCD-like $SU(N)_D$ dark sector, but can also give rise to a chiral $SU(N)_D$ dark sector, such as the quark content in eq.~\eqref{chiralfermion1}. The chiral strongly coupled dark sectors of this type are less studied due to the limited understanding of chiral gauge dynamics. Nonetheless, this also suggests the possibility of rich new physics and novel insights into the yet unexplored dark sector!

\section{Running couplings and confinement scales at the two-loop level}\label{App:Twoloop}

Since the comparability of the running couplings and confinement scales is central to the model, we present a two-loop analysis to validate this feature in this appendix. For the coupling constant $\alpha_i$ of the gauge group $G_i\times G_j$, the two-loop beta functions are given by 
\begin{align}
\frac{d\,\alpha_i^{-1}}{d\,\text{ln}\,\mu}=\frac{1}{2\pi}\,\left(b_i+\frac{\alpha_i}{4\pi}c_{ii}+\frac{\alpha_j}{4\pi}c_{ij}\right)~, 
\end{align}
where the one-loop coefficient $b_{i}$ is given in eq.~\eqref{1loop}. Two terms appear at the two-loop level: one arises from the contributions of the particles charged under the gauge group $G$,
\begin{align}
c_{ii}=\frac{34}{3}C^2_2(G)
-\left(\frac{10}{3}C_2(G)+2C_2(R_f)\right)n_fT(R_f)
-\left(\frac{1}{3}C_2(G)+2C_2(R_s)\right)n_sT(R_s),
\end{align}
and the other originates from contributions of the particles that are also charged under the other gauge group $G'$, given by
\begin{align}
c_{ij}=-2C_2(R_f)n'_fT(R_f)-2C_2(R_s)n'_sT(R_s),
\end{align}
where $n'_f$ and $n'_s$ denote the number of active bicharged Weyl fermions and real scalars, respectively, which play the role of connecting the two running coupling constants. The beta function of $\alpha_j$ can be straightforwardly derived by interchanging $(i\leftrightarrow j)$. With all these equations, we can write down the coupled beta functions in our model as
\begin{align}\label{twoloopODE}
\frac{d\,\alpha_D^{-1}}{d\,\text{ln}\,\mu}=\frac{1}{2\pi}\,\left(b_D+\frac{\alpha_D}{4\pi}c_{DD}+\frac{\alpha_s}{4\pi}c_{DS}\right),~~
\frac{d\,\alpha_s^{-1}}{d\,\text{ln}\,\mu}=\frac{1}{2\pi}\,\left(b_s+\frac{\alpha_s}{4\pi}c_{ss}+\frac{\alpha_D}{4\pi}c_{sD}\right),
\end{align}
where the values of the coefficients are summarized in table~\ref{tab:beta2}. 

\begin{table}[t]
\centering
\begin{tabular}{|c|c|c|c|c|c|c|}
\hline
coefficient  & pure gauge & [$1$, $m_t$] & [$m_t$, $10^5$] & [$10^5$, $10^7$]  & [$10^7$, $10^{11}$] & [$10^{11}$, $10^{15}$]  \\ \hline
$c_{ss}$   & $102$  & $64-38\sfrac{2}{3}$  & $26$  & $-100$  & $-150\sfrac{2}{3}$ & $-201\sfrac{1}{3}$ \\ \hline
$c_{sD}$   & $0$  & $0$      & $0$  & $0$  & $-10\sfrac{2}{3}$ & $-21\sfrac{1}{3}$ \\ \hline
$c_{DD}$   & $102$  & $75\sfrac{1}{2}$   & $63$  & $-75\sfrac{1}{2}$ & $-152$ & $-214\sfrac{1}{2}$ \\ \hline
$c_{Ds}$    & $0$  & $0$   & $0$  & $0$ & $-7\sfrac{1}{2}$ & $-15$ \\ \hline
$\Delta c$    & $0$  & $11\sfrac{1}{2}-36\sfrac{5}{6}$   & $37$  & $24\sfrac{1}{2}$ & $1\sfrac{5}{6}$ & $-6\sfrac{5}{6}$ \\ \hline
\end{tabular}
\caption{The two-loop beta function coefficients of QCD and DC across different intervals of scales in units of GeV. For comparison, we also present the quantity $\Delta c \equiv (c_{DD}+c_{Ds})-(c_{ss}+c_{sD})$, which provides a rough estimate of the difference between the two beta functions. \label{tab:beta2}}
\end{table}

\begin{figure}[tbp]
\centering
\includegraphics[width=0.83\textwidth]{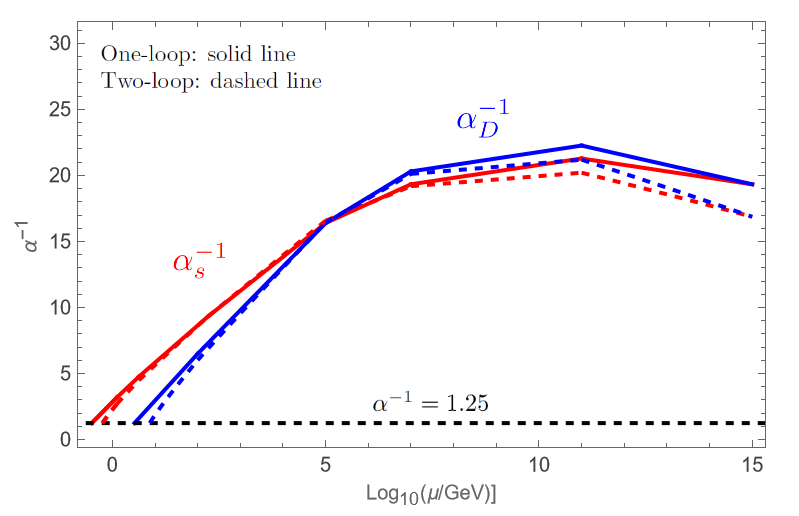}
\caption{\label{fig:twoloop} One-loop (solid lines) and two-loop (dashed lines) running of ${\alpha_s}$ (red) and ${\alpha_D}$ (blue), assuming ${\alpha'_4}=\infty$. A realistic value of ${\alpha'_4}$ would shift the blue curve upward as shown in figure \ref{fig:alpha4}.}
\end{figure}

Besides the contributions from the gauge sector, we also consider the two-loop contributions arising from the scalar sector due to the presence of sizable Yukawa couplings. To simplify the analysis, we retain only the top Yukawa coupling and its one-loop running, while neglecting the contributions from other light quarks due to their small Yukawa couplings. An additional ${\alpha_t}/{2\pi}$ contribution appears inside the bracket of eq.~\eqref{twoloopODE} above the top mass threshold. Similarly, there are also Yukawa couplings in the dark sector between the dark Dirac fermion and the scalar corresponding to the radial mode of the VEV $\langle\Phi'\rangle$. Again, we only include the heaviest dark Dirac fermion, which has a Yukawa coupling $y_{f,3}\sim\mathcal{O}(1)$, giving rise to an additional contribution of ${\alpha_y}/{2\pi}$. Unlike the top Yukawa coupling, this contribution only becomes active above the scale $m_{f,3}=10^5$ GeV. The final numerical result is shown in figure~\ref{fig:twoloop}.

Following the same criterion that $\alpha^{-1}(\Lambda_{\rm confinement}) = 1.25$, the QCD confinement scale obtained from the two-loop calculation is $\Lambda_{\rm QCD} = 0.59~\text{GeV}$, while the DC confinement scale becomes $\Lambda_{\rm DC} = 7.5~\text{GeV}$. Note that the larger value of the confinement scale does not imply a modification of SM QCD, but merely serves as a reference value under this criterion. The physically meaningful quantity is the ratio $\Lambda_{\rm DC}/\Lambda_{\rm QCD}$, which is proportional to $m_D/m_B$ up to an $\mathcal{O}(1)$ factor. In this appendix, we specifically aim to examine the change in this ratio after including the two-loop effects, obtaining
\begin{align}
\frac{\Lambda_{\rm DC}}{\Lambda_{\rm QCD}}\Big|_{\rm one-loop}=10.0~~\to\quad
\frac{\Lambda_{\rm DC}}{\Lambda_{\rm QCD}}\Big|_{\rm two-loop}=12.7~~.
\end{align}
The numerical result shows good agreement, with only an $\mathcal{O}(10\%)$ difference, indicating that higher-loop contributions do not significantly alter the result. Examining the details, we find that above $10^5$~GeV, despite large negative two-loop contributions for both $\alpha's$, the difference between the two running couplings is small, because the couplings are relatively weak ($\alpha/4\pi\sim 1/200$) and the two-loop coefficients are comparable (see Table~\ref{tab:beta2}). Below $10^5$ GeV, the larger two-loop contribution to the running of $\alpha_D$ drives it more rapidly toward the confinement scale compared to $\alpha_s$, resulting in a larger ratio. The desired dark color confinement scale can still be achieved when a finite $\alpha'_4$ is taken into account, leaving the benchmark model and dark sector phenomenology unchanged.


\bibliographystyle{jhepbst}
\bibliography{DM_Ref}{}

\end{document}